\documentclass[aps,prd,12pt,nofootinbib,superscriptaddress]{revtex4-2}
\usepackage{epsfig}
\usepackage{graphicx}
\usepackage{amsmath}
\usepackage{amssymb}
\usepackage{mathrsfs}
\usepackage{verbatim}
\usepackage{dutchcal}
\usepackage{comment}
\usepackage{float}
\usepackage[normalem]{ulem}
\usepackage{xcolor}
\usepackage{slashed}

\newcounter{fig}

\newcommand{\bea}{\begin{eqnarray}}
\newcommand{\eea}{\end{eqnarray}}
\newcommand{\be}{\begin{equation}}
\newcommand{\ee}{\end{equation}}

\def\({\left(}
\def\){\right)}

\newcommand{\re}[1]{(\ref{#1})}

\newcommand{\tr}{\mbox{Tr}}

\def\rlx{\relax\leavevmode}
\def\IR{\rlx\hbox{\rm I\kern-.18em R}}
\def\one{\hbox{{1}\kern-.25em\hbox{l}}}

\newcommand{\eqn}{\begin{eqnarray}}
\newcommand{\eqnx}{\end{eqnarray}}

\date{\today}

\begin{document}

\title{Gravitating Skyrmions with localized fermions
}

\author{
Vladimir Dzhunushaliev
}
\email{v.dzhunushaliev@gmail.com}
\affiliation{
Department of Theoretical and Nuclear Physics,  Al-Farabi Kazakh National University, Almaty 050040, Kazakhstan
}
\affiliation{
Institute of Nuclear Physics, Almaty 050032, Kazakhstan
}
\affiliation{
Academician J.~Jeenbaev Institute of Physics of the NAS of the Kyrgyz Republic, 265 a, Chui Street, Bishkek 720071, Kyrgyzstan
}
\affiliation{
International Laboratory for Theoretical Cosmology, Tomsk State University of Control
Systems and Radioelectronics (TUSUR),
Tomsk 634050, Russia
}

\author{Vladimir Folomeev}
\email{vfolomeev@mail.ru}
\affiliation{
Institute of Nuclear Physics, Almaty 050032, Kazakhstan
}
\affiliation{
Academician J.~Jeenbaev Institute of Physics of the NAS of the Kyrgyz Republic,
265 a, Chui Street, Bishkek 720071, Kyrgyzstan
}
\affiliation{
International Laboratory for Theoretical Cosmology, Tomsk State University of Control
Systems and Radioelectronics (TUSUR),
Tomsk 634050, Russia
}

\author{Jutta Kunz}
\email{jutta.kunz@uni-oldenburg.de}
\affiliation{Institute of Physics,
Carl von Ossietzky University Oldenburg, Germany
Oldenburg D-26111, Germany}

\author{Yakov Shnir}
\email{shnir@theor.jinr.ru}
\affiliation{BLTP, JINR, Dubna 141980, Moscow Region, Russia}
\affiliation{Institute of Physics,
Carl von Ossietzky University Oldenburg, Germany
Oldenburg D-26111, Germany}
\affiliation{Hanse-Wissenschaftskolleg, Lehmkuhlenbusch 4, 27733 Delmenhorst, Germany
}
\date{\today}

\begin{abstract}
We consider self-gravitating Skyrmions in the presence of Dirac fermions, that carry spin and isospin.
By varying the gravitational and the Yukawa coupling constants, we investigate the spectral flow of the fermion eigenvalue associated with a zero mode in the absence of gravity.
We demonstrate that the backreaction of the fermion can strongly influence the Skyrmion-fermion configurations.
In particular, the energy conditions may be violated, and regular anti-gravitating asymptotically flat solutions with negative ADM mass may emerge.
\end{abstract}

\date{\today}

\maketitle
\newpage
\section{Introduction}
For a long time now, much attention has been paid to soliton solutions of various classical field theories.
These are regular localized field configurations with finite energy.
Well known examples of such topological solitons in (3+1)-dimensions
are monopoles in the Yang-Mills-Higgs model \cite{tHooft:1974kcl,Polyakov:1974ek},
Skyrmions \cite{Skyrme:1961vq,Skyrme:1962vh} and Hopfions \cite{Faddeev:1975tz,Faddeev:1996zj}
(for reviews, see, for example, Refs.~\cite{Manton:2004tk,Shnir:2018yzp}).
In particular, the groundbreaking work by Skyrme \cite{Skyrme:1961vq,Skyrme:1962vh}
gave rise to an increasing interest in the study of various aspects of such solutions
in a wide variety of physical systems, both in flat and curved spacetime.

Flat space topological solitons can be minimally coupled to gravity.
The emerging systems of the Einstein-matter field equations may then support the
existence of gravitating spatially localized globally regular solutions.
Endowing these gravitating soliton solutions with an event horizon, the so-called hairy black holes arise,
which possess nontrivial matter fields outside their event horizon, i.e., they carry hair
(see, e.g., Refs.~\cite{Volkov:1998cc,Volkov:2016ehx}).
Well known examples include black holes within monopoles \cite{Lee:1991vy,Breitenlohner:1991aa,Breitenlohner:1994di}
and Skyrmions \cite{Luckock:1986tr,Glendenning:1988qy,Droz:1991cx,Heusler:1991xx,Bizon:1992gb,Heusler:1993ci}.

A notable feature of topological solitons is the remarkable relation between the topological charge of the
configuration and the fermionic zero modes localized on the soliton, provided by
the fundamental Atiyah-Patodi-Singer index theorem \cite{Atiyah:1975jf}.
In particular, in the background of a soliton of topological degree one 
the fermion spectrum exhibits a spectral flow of the eigenvalues with a normalizable bound mode crossing zero.

Study of the localization of fermionic modes on solitons started with the pioneering work of fermion bound states on a vortex line \cite{Caroli:1964},
and was followed by the investigation of fermion bound states on kinks
\cite{Jackiw:1975fn,Dashen:1974cj,Chu:2007xh}, 
sphalerons \cite{Dashen:1974cj,Nohl:1975jg,Boguta:1985ut},
monopoles \cite{Callias:1977cc},
Skyrmions \cite{Hiller:1986ry,Kahana:1984be,Balachandran:1998zq},
and domain walls \cite{Stojkovic:2000ub}.
Highlights of those and related studies include 
the appearance of fractional fermion number of solitons \cite{Jackiw:1975fn,Jackiw:1981ee}, 
monopole catalysis of proton decay \cite{Rubakov:1982fp,Callan:1982au},
or the emergence of superconducting cosmic strings \cite{Witten:1984eb}.

The nature of the fermionic zero modes depend on the associated solitons.
For instance, for monopoles one finds an exact zero mode localized
on them, irrespective of the strength of the Yukawa coupling \cite{Jackiw:1975fn}.
For Skyrmions, in contrast, one observes spectral flow of the eigenvalues
\cite{Hiller:1986ry,Kahana:1984be,Balachandran:1998zq}. 
The fermion eigenvalue then depends crucially on the value of the Yukawa coupling,
emerging from the positive continuum at a critical value of the coupling strength, then crossing zero at some other particular value and finally approaching the negative continuum.

While it is natural to first explore the properties of fermions in the background
of the solitons, a subsequent investigation of the backreaction of the fermions
on the solitons is called for, in particular, when the fermion-soliton coupling is no longer weak.
In such a case, the properties of the solitons may be significantly affected.
Examples range from the distortion of domain walls \cite{Gani:2010pv}, kinks 
\cite{Amado:2014waa,Klimashonok:2019iya}, Skyrmions \cite{Perapechka:2018yux}, and $CP^2$ solitons \cite{Amari:2024rpm},
via the emergence of soliton bound states 
\cite{Perapechka:2019upv,Perapechka:2019vqv},
to the outcome of soliton collisions \cite{Campos:2022flw,Gani:2022ity,Bazeia:2022yyv},
where the inclusion of the Dirac sea may possibly mitigate the effects somewhat \cite{Saadatmand:2022htx,Weigel:2023fxe}.

When including gravity, one needs to take into account the fundamental quantum character of the fermions \cite{Taub:1937zz}.
This is typically done by describing the Dirac field with normalized wave functions, and by imposing the Pauli principle via the appropriate particle numbers of the considered states.
The basic approximation scheme employed in the literature then consists of 
(i) considering only single-particle fermion states, 
(ii) ignoring second quantization of the fields, 
(iii) treating gravity only classically. 

Based on this approximation scheme, for instance, fermions were studied 
in the background of the Einstein-Yang-Mills sphaleron barrier and the
level crossing phenomenon was observed \cite{Volkov:1994tp},
with a fermionic zero mode at the top of the barrier,
very much in analogy with the electroweak case \cite{Kunz:1993ir}.
The dynamical evolution of the fermionic zero mode localized on the self-gravitating $SU(2)$ monopole was studied in Ref.~\cite{Dzhunushaliev:2023ylf}.
Considering only the Einstein-Dirac equations, 
the existence of regular localized configurations was shown in Ref.~\cite{Finster:1998ws}.
In fact, such configurations, the so-called Dirac stars, share a number of properties
with boson stars
\cite{Herdeiro:2019mbz,Dzhunushaliev:2018jhj,Dzhunushaliev:2019kiy,Herdeiro:2021jgc}.
Interestingly, self-gravitating fermions may affect the metric dramatically, leading to significant violations of the energy conditions and allowing for (traversable) wormholes \cite{Blazquez-Salcedo:2019uqq,Blazquez-Salcedo:2020czn,Bolokhov:2021fil,Konoplya:2021hsm}.
Violation of the energy conditions by spinor fields was also seen to give rise to 
notable effects in cosmology
\cite{Armendariz-Picon:2003wfx,Cai:2008gk}. 

Here we consider in detail the spectral flow of the gravitating Skyrmion-fermion system~\cite{Dzhunushaliev:2024iag},
where the spherically symmetric 
fermionic mode is occupied by a single fermion,
that backreacts on the gravitating Skyrmion.
Without the fermion, the gravitating Skyrmion exhibits two branches of solutions.
The lower mass Skyrmion branch starts from the flat space soliton, 
and merges with the higher mass Bartnik-McKinnon (BM) 
branch at a maximal value of the gravitational coupling constant
\cite{Glendenning:1988qy,Bizon:1992gb,Heusler:1991xx,Heusler:1993ci,Bartnik:1988am}).
Since in flat space the Skyrmion-fermion system exhibits spectral flow, 
when the Yukawa coupling is varied, 
it is interesting to investigate the effect of gravity
on the spectral flow of the backreacting Skyrmion-fermion system.

The paper is organized as follows:
We start with a brief description of the theoretical setting in section 2, 
where we present the action, the Ansatz for the fields, the resulting set of
coupled equations, and the boundary conditions.
In section 3 we discuss the numerical approach.
Subsequently we give a detailed discussion of the numerical results.
We end with our conclusions in section 4.

\section{Theoretical Setting}

\subsection{Einstein-Skyrme-Dirac Action}

We consider the Einstein-Skyrme-Dirac system in (3+1)-dimensions,
consisting of the Einstein-Hilbert action with curvature scalar $R$ 
and Newton's constant $G$ \footnote{We use natural units with $c=\hbar=1$ throughout the paper.}
coupled minimally to the matter fields, represented 
by the matter Lagrangian ${\cal L}_m$,
\be
S = \int d^4 x \sqrt{-\cal g}
\left(-\frac{R}{16 \pi G} + {\cal L}_m \right)\ .
\label{Lag-tot}
\ee
The matter Lagrangian ${\cal L}_m$ itself consists of three parts,
\be
{\cal L}_m= {\cal L}_{\text{Sk}}+{\cal L}_{\text{sp}} + {\cal L}_{\text{int}} \ .\nonumber
\ee

The first part of ${\cal L}_m$ is the Skyrme Lagrangian, minimally coupled to gravity
\be
{\cal L}_{\text{Sk}}=- \frac{f_\pi^2}{4}\,\,\tr \(\partial_\mu U\,\partial^\mu U^\dagger\)+
\frac{1}{32\,a_0^2} \,\tr \(\left[\partial_\mu U\, U^\dagger,\, \partial_\nu U\,U^\dagger\right]^2\)\,.
\label{SkLag}
\ee
It is expressed in terms of the matrix-valued Skyrme field $U\in SU(2)$ \cite{Skyrme:1961vq,Skyrme:1962vh}, 
\be
U=\phi_0\,\mathbb{I}+i\,\sum_{n=1}^3\phi_n\,\tau_n \ ,  \label{decomposition}
\nonumber
\ee
where $U$ is composed of a scalar component $\phi_0$ and the pseudoscalar isospin triplet field $\phi_k$,
and $\tau_n$ denotes the Pauli matrices.
The four chiral field components $\phi_a = (\phi_0,\phi_k)$ need to satisfy the sigma-model constraint $\phi_a \cdot \phi^a = 1$.
We note that we do not consider a mass term for the chiral fields.
The two parameters of the model $f_\pi$ and $a_0$ have dimensions $[f_\pi] =L^{-1}$ and $[a_0]= L^0$, respectively.
The Skyrme field is required to attain its vacuum value, the identity, at spatial infinity, $U \xrightarrow[\vec r \to\infty]{} {\mathbb{I}}$.
It is a map $U:S^3 \mapsto S^3$ characterized by an integer-valued winding number \cite{Skyrme:1961vq,Skyrme:1962vh}.

The second part of ${\cal L}_m$ is the Dirac Lagrangian for the isospinor fermions $\psi$, 
\be
{\cal L}_{\text{sp}} =
 \frac{\imath}{2}\left[
(\hat {\slashed{D}}\bar \psi) \psi - \bar \psi \hat {\slashed D} \psi \right] - m\bar \psi \psi \, ,
\label{Lag_ferm}
\ee
where the spinor covariant derivative $\hat {\slashed D} = \gamma^\mu \hat D_\mu$
contains the Dirac matrices $\gamma^\mu$ in the standard representation in a curved spacetime, and 
\be
\hat D_\mu \psi = (\partial_\mu - \Gamma_\mu )\psi \ 
\nonumber
\ee
with the spin connection matrices $\Gamma_\mu$ \cite{Eguchi:1980jx,Dolan:2015eua}.
In (\ref{Lag_ferm}) we have also added a bare mass $m$ for the fermions.

The last part of ${\cal L}_m$ describes the chiral Skyrmion-fermion interaction 
\be
{\cal L}_{\text{int}}=h \bar \psi \, U^{\gamma_5}\, \psi, \qquad U^{\gamma_5} \equiv \frac{\mathbb{I} +  \gamma_5}{2}U
+ \frac{\mathbb{I} -  \gamma_5}{2}U^\dagger\, ,
\label{L_int}
\ee
where the Yukawa coupling constant $h$ determines the strength of the interaction,
and $U^{\gamma_5}$ incorporates the pseudoscalar pion-fermion coupling,
realized with help of the Dirac matrix $ \gamma^5$, that corresponds to the respective Dirac matrix in curved spacetime.
We note that this matrix is the same in both flat and curved spacetime,
\begin{equation}
\begin{split}
    {\gamma}^5 &= -\frac{\imath}{4!} E_{\alpha \beta \rho \sigma}
    \gamma^\alpha \gamma^\beta \gamma^\rho \gamma^\sigma =
    -\frac{\imath}{4!} \sqrt{-\cal g} \epsilon_{\alpha \beta \rho \sigma} e_a^\alpha e_b^\beta e_c^\rho e_d^\sigma
    \hat \gamma^a \hat \gamma^b \hat \gamma^c \hat \gamma^d \\
    &=
    \frac{\sqrt{-\cal g}}{4!}  \left(
    \epsilon_{\alpha \beta \rho \sigma} \epsilon^{a b c d} e_a^\alpha e_b^\beta e_c^\rho e_d^\sigma
    \right) \hat \gamma^5 = \hat \gamma^5 \, , 
    \label{gamma5}
\end{split}
\nonumber
\end{equation}
where $E_{\alpha \beta \rho \sigma} = \sqrt{-\cal g} \epsilon_{\alpha \beta \rho \sigma} $ is the Levi-Civita tensor in curved spacetime,
and $\hat \gamma^5=-\imath \hat \gamma^0 \hat \gamma^1 \hat \gamma^2 \hat \gamma^3$ is the Dirac matrix in Minkowski spacetime
(i.e., $\hat \gamma^a$ denotes the usual flat spacetime Dirac matrices).

We now transform to dimensionless quantities.
We therefore introduce 
the dimensionless radial coordinate $\tilde r = a_0 f_\pi  r$,
the dimensionless Dirac field $\tilde \psi = \psi/\sqrt{a_0 f_\pi^3 }$,
the dimensionless bare fermion mass $\tilde m = m/(a_0 f_\pi)$,
the dimensionless Yukawa coupling constant $\tilde h = h/(a_0 f_\pi)$,
and the effective gravitational coupling $\alpha^2=4\pi G f_\pi^2$.
We also introduce the dimensionless ADM mass 
$\tilde M = \alpha^2 M/(4\pi f_\pi/a_0)$
(for its explicit definition see Eq.~\re{expres_mass} below). 
Hereafter we drop the tilde in the dimensionless quantities for the sake of brevity.

In terms of dimensionless quantities the Skyrme Lagrangian \re{SkLag} then reads in component notation 
\be
\label{Lag}
{\cal L}_{\text{Sk}}=
 \partial_\mu \phi^a \partial^\mu \phi^a
-\frac12 (\partial_\mu \phi^a \partial^\mu \phi^a)^2 +
\frac{1}{2} (\partial_\mu \phi^a \partial_\nu \phi^a)(\partial^\mu \phi^b \partial^\nu \phi^b)\, ,
\nonumber
\ee
and the interaction Lagrangian \re{L_int} takes the form \cite{Gell-Mann:1960mvl,Hiller:1986ry,Kahana:1984be,Balachandran:1998zq,Krusch:2003xh,Shnir:2002dw}
\be
{\cal L}_{\text{int}}=h \bar \psi [\phi_0 + i \gamma_5(\phi^a \cdot \sigma^a)] \psi \, .
\label{L_int-comp}
\nonumber
\ee

The corresponding components of the stress-energy tensor are
\be
T^{\mu\nu}=T^{\mu\nu}_{\text{Sk}} + T^{\mu\nu}_{\text{sp}} ,
\label{SET_tot}
\ee
where the stress-energy tensor of the gravitating Skyrmion is
\be
T^{\mu\nu}_{\text{Sk}}=2\,\left[ \partial^\mu \phi_a\,\partial^{\nu}\phi^a-
\left(\partial^{[\mu} \phi^a\,\partial^{\alpha]} \phi^b\right)\,\left(\partial^{[\nu} \phi_a
\partial_{\alpha]}\phi_b\right)\right]
-g^{\mu\nu}\,\left[\left( \partial_\alpha \phi_a \right)^2-\frac12 \left(
\partial_{[\alpha} \phi_a \,\partial_{\beta]} \phi_b \right)^2
\right]\, ,
\label{T-Sk}
\ee
and the stress-energy tensor of the gravitating isospinor is
\be
T^{\mu\nu}_{\text{sp}}=
\frac{\imath}{4}\left[
        \bar \psi  \gamma^{\mu} (\hat D^\nu \psi)
        + \bar \psi \gamma^\nu (\hat D^\mu \psi)
        - (\hat D^\mu \bar \psi) \gamma^{\nu } \psi
        - (\hat D^\nu \bar \psi) \gamma^\mu\psi
    \right] - g^{\mu \nu} {\cal L}_{\text{sp}} \, .
\label{T-sp}
\ee

\subsection{Spherically symmetric Ansatz and field equations}

Here we focus on spherically symmetric solutions of 
the equations obtained from
the action \re{Lag-tot},
that are based on the above set of assumptions.
Treating the gravitational field on a purely classical level,
we employ Schwarzschild-like coordinates 
and a static spherically symmetric metric Ansatz, analogous to the one 
used for gravitating Skyrmions (see, e.g., Refs.
\cite{Glendenning:1988qy,Bizon:1992gb,Heusler:1991xx,Heusler:1993ci}),
\begin{equation}
    ds^2 = \sigma^2 (r) N(r) dt^2 - \frac{dr^2}{N(r)} - r^2 \left(
        d \theta^2 + \sin^2 \theta d \varphi^2
    \right) \ .
\label{metrics}
\end{equation}

This metric Ansatz implies the following form of the orthonormal tetrad:
\be
    e^a_{\phantom{a} \mu} = \text{diag} \left\lbrace
        \sigma \sqrt{N}, \frac{1}{\sqrt{N}}, r, r \sin \theta
    \right\rbrace ,
\label{tetrad}
\ee
such that $ds^2=\eta_{ab}(e^a_\mu dx^\mu )(e^b_\nu dx^\nu)$, where $\eta_{ab}=(+1,-1,-1,-1)$ is the Minkowski metric.
We recall that the Dirac matrices in curved spacetime are given by 
$\gamma^\mu= e^\mu_{\phantom{a} a} \hat \gamma^a$ 
with $\hat \gamma^a$ being the usual flat spacetime Dirac matrices.

Considering the scalar fields and the resulting solitons
also on a purely classical level, we adopt the 
usual hedgehog Ansatz for a Skyrmion of topological charge one,
\be
 U = \cos \left( F(r)\right)  \mathbb{I} + \imath \sin \left( F(r)\right)  \left(
        \phi_a n^a\right) \, ,
\label{hedgehog}
\ee
with radial unit vector $n^a = \{\sin (\theta ) \cos (\varphi ),\sin (\theta ) \sin (\varphi ),\cos (\theta )\}$. 

Based on the above set of assumptions, we treat the Dirac field 
in terms of a quantum wave function, that is normalized.
Since we consider an isospinor-spinor field, we then couple spin and isospin
to zero, to obtain a singlet state.
In order to respect the Pauli principle, 
and thus the fermionic nature of the Dirac field,
we occupy this singlet state only with a single fermion.
Thus, we here consider the spectral flow only for this singlet state,
that is backreacting on the gravitating Skyrmion (with topological
number one).

The appropriate Ansatz for the isospin-spin singlet Dirac field then features
an Ansatz with a harmonic time dependence,
and can be expressed in terms of two $2\times 2$ matrices $\chi$ and $\eta$ \cite{Jackiw:1975fn,Jackiw:1976xx},
\be
\psi =  e^{-\imath\omega t}\begin{pmatrix}
        \chi \\
        \eta
    \end{pmatrix}
\quad \text{with} \quad
 \chi =
    \frac{u(r)}{\sqrt{2}} \begin{pmatrix}
        0   &   -1 \\
        1   &   0
    \end{pmatrix}, \quad
  \eta =
     \imath \, \frac{v(r)}{\sqrt{2}} \begin{pmatrix}
        \sin \theta e^{- \imath \varphi}    &   - \cos \theta \\
        - \cos \theta    &   -\sin \theta e^{\imath \varphi}
    \end{pmatrix} .
\label{spherical-fermions}
\ee
Here $\omega$ denotes the eigenvalue of the Dirac operator,
and the real functions $u(r)$ and $v(r)$ depend only on the radial coordinate.

Substitution of the Ans\"atze \re{metrics}, \re{hedgehog}, 
and \re{spherical-fermions} into the general action \re{Lag-tot} yields the reduced Einstein-Hilbert Lagrangian
\be
{\cal L}_{\text{GR}} = - \frac{1}{2\alpha^2 } \left( \frac{N'}{r} + \frac{N - 1}{r^2} \right) ,
\label{LagrGR}
\ee
where we denote the radial derivative with a prime,
the Skyrme Lagrangian 
\be
{\cal L}_{\text{Sk}}=-\frac12 \left[N (F^\prime)^2 + \frac{2 \sin^2 F}{r^2} \right] - \frac{\sin^2 F}{r^2} \left[N (F^\prime)^2
+ \frac{\sin^2 F}{2r^2} \right] \, ,
\label{LagrSk}
\ee
the fermion Lagrangian 
\be
{\cal L}_{\text{sp}} =  \sqrt{N} \left( v u' - u v' \right)
    - \frac{2 u v}{r}
    + \frac{ \omega  \left(u^2 + v^2\right)}{\sigma \sqrt{N} } +
     m  \left(-u^2 + v^2\right)\,  ,
\label{LagrD}
\ee
and the interaction Lagrangian 
\be
{\cal L}_{\text{int}} = h\left[\cos F (u^2 - v^2) - 2  u v \sin F\right] \, .
\label{Lagrint}
\ee
We note that in the expressions \re{LagrGR}~-~\re{Lagrint} 
all particular Lagrangians 
have been expressed in dimensionless quantities (corresponding to a rescaling 
of the total Lagrangian by the factor $1/(a_0^2 f_\pi^4)$).

The dimensionless total Lagrangian then consists of the Lagrangians \re{LagrGR} - \re{Lagrint}.
Multiplying this total Lagrangian by $\sqrt{-\cal g}$
and varying the resulting expression with respect to the functions $N$, $\sigma$, $F$, $u$, and $v$, we obtain the following set of five coupled ordinary differential equations:
\begin{align}
    N' + \frac{1}{r}\left(N-1\right) = & - \alpha^2 \left[
      \frac{N \left(2 \sin^2 F + r^2 \right)}{r}(F')^2
     + \frac{\sin^2 F}{r}\left(2 + \frac{\sin^2 F}{r^2}\right)
    + 2 \omega \frac{r \left( u^2 +  v^2\right)}{\sigma \sqrt{N}}
    \right] ,
\label{GR1}\\
    \frac{\sigma'}{\sigma} = & - \alpha^2  \Big[
        -  \frac{2 \sin^2 F + r^2}{r} (F')^2+
    h r \frac{\left(v^2- u^2\right)\cos F+2u  v \sin F }{N}
\nonumber \\
    &
    -2  \omega
    \frac{r\left( u^2 +  v^2\right)}{\sigma N^{3/2}}
    + 2 \frac{u  v}{N}
    + m \frac{r\left( u^2 -  v^2\right)}{N}
    \Big] ,
\label{GR2}
\end{align}
\begin{equation}
\begin{split}
    &
     F'' +   \frac{ \sin (2 F)}{2 \sin^2 F + r^2}(F')^2
     +  \left(
        \frac{2 r}{2 \sin^2 F + r^2}
        + \frac{N'}{N} + \frac{\sigma '}{\sigma }
    \right)F'
    \\
    &
    +  h r^2 \frac{\left( v^2- u^2\right)\sin F - 2 u v \cos F}
    {N \left(2 \sin^2 F + r^2\right)} -
    \frac{\sin^2 F + r^2}{r^2 N \left(2 \sin^2 F  + r^2\right)} \sin(2 F)
    = 0,
\end{split}
\label{Skyrma_eqn}
\end{equation}
\begin{align}
     u' +   \left[
         \frac{N'}{4 N} + \frac{\sigma '}{2 \sigma }
            + \frac{1}{r}\left(1-\frac{1}{\sqrt{N}} \right)
    \right] u
    -\frac{ h}{\sqrt{N}}\left( u \sin F+   v \cos F\right) + \omega\frac{ v}{N \sigma }+m\frac{ v}{\sqrt{N}}
    = & 0,
\label{Dirac_eq1}  \\
     v' +   \left[
         \frac{N'}{4 N} + \frac{\sigma '}{2 \sigma }
            + \frac{1}{r}\left(1+\frac{1}{\sqrt{N}} \right)
    \right] v
    +\frac{h}{\sqrt{N}}\left( v \sin F-
     u \cos F\right) -  \omega\frac{u}{N \sigma }+ m \frac{ u}{\sqrt{N}}
    = & 0 .
\label{Dirac_eq2}
\end{align}

The system of equations \eqref{GR1} - \eqref{Dirac_eq2} is supplemented by the normalization condition of the localized fermionic mode,
\begin{equation}
  \int dV\, \psi^\dagger \psi =
\frac{4\pi}{a_0^2} \int_{0}^{\infty}\frac{u^2 + v^2}{\sqrt{N}}r^2 dr
=    1 .
\label{norm}
\end{equation}

\subsection{Asymptotic behavior and boundary conditions}

To obtain an appropriate set of boundary conditions, 
an asymptotic expansion of the five unknown functions is made 
for the boundaries $r \to 0$ and $r \to \infty $ 
and inserted into the set of the field equations~\re{GR1} - \re{Dirac_eq2}.
Then regularity and asymptotic flatness of the solutions are imposed together
with the demand that the Skyrmion profile function $F(r)$ 
should correspond to a configuration of topological charge one.

The expansions at the origin then read
\begin{equation}
N \approx 1 + \frac{1}{2} N_2 r^2, \quad
\sigma \approx \sigma_0 + \frac{1}{2} \sigma_2 r^2, \quad
F \approx \pi +F_1 r, \quad
u \approx  u_0+\frac{1}{2} u_2 r^2, \quad
v \approx v_1 r \, ,
\label{expand}
\nonumber
\end{equation}
where the parameters $\sigma_0$, $F_1$, and $u_0$ are found numerically,
while the expansion coefficients $N_2$, $\sigma_2$, $u_2$, and $v_1$ are
obtained from Eqs.~\re{GR1} - \re{Dirac_eq2} in terms of the above parameters.

Asymptotic flatness of the spacetime implies that the metric~\re{metrics} approaches the Minkowski metric at spatial infinity.
The asymptotic behavior of the metric and matter field functions 
is then mostly given by
\be
\begin{split}
     N & \approx 1 - \frac{2 M}{r} , \quad
     \sigma \approx 1 - \frac{\alpha^2 F_2^2}{r^4},\quad
     F \approx  \frac{F_2}{r^2}, \quad \\
      u & \approx v_2 \sqrt{\frac{h - \omega}{h + \omega}}
                    e^{- r \sqrt{h^2 - \omega^2}}, \quad
      v \approx  v_2 e^{- r \sqrt{h^2 - \omega^2}}, 
    \label{asympt}
\end{split}
\ee
where $M$, $F_2$, and $v_2$ are integration constants,
with $M$ corresponding to the dimensionless mass of the system.
We note that for localized solutions with vanishing $M$ obtained below
the asymptotic behavior of the metric function $N$ changes to
\be
    N \approx 1 + \frac{\alpha^2 F_2^2}{r^4}  ,
\label{asympt_zero_mass}
\nonumber
\ee
while the asymptotic behavior of the metric function $\sigma$, the Skyrme field function $F$, and the spinor components $u$ and $v$ remains the same.
Clearly, for localized fermionic modes to exist,
it is required that $|h|>|\omega|$.

Explicitly, we impose the following sets of boundary conditions 
at the origin and at infinity,
\be
\begin{split}
N(0)=&1, \quad \partial_r \sigma(0)=0, \quad F(0)=\pi, 
\quad \partial_r u(0)=0, \quad v(0)=0 \, ;\\
N(\infty)=&1, \quad \sigma(\infty)=1, \quad F(\infty)=0,
\quad u(\infty)=0, \quad v(\infty)=0  \, . 
\label{BC}
\end{split}
\ee

\section{Numerical methods and results}

\subsection{Numerical approach}

We solve the system of mixed order differential equations 
\re{GR1} - \re{Dirac_eq2} together with the constraint equation 
imposed by the normalization condition \re{norm} 
subject to the above set of boundary conditions 
\re{BC}.
The parameters entering the equations are the effective gravitational
coupling $\alpha$, the Yukawa coupling $h$, and, in principle,
the dimensionless Skyrme parameter $a_0$ and the bare fermion mass $m$.
While we vary $\alpha$ and $h$ in our numerical calculations,
we fix the Skyrme parameter $a_0$ to a particular
value\footnote{In the context of applications of the Skyrme model 
as a model of nuclear physics, the usual choice is 
$a_0=4.84$~\cite{Adkins:1983ya}, although other values were 
also considered, see, e.g., Ref.~\cite{Manton:2006tq}.}, $a_0=9.633$.
Furthermore, we set the bare fermion mass $m$ to zero, 
unless explicitly stated.

In order to map the semi-infinite range of the radial variable $r$ 
to the finite interval $[0, 1]$, we introduce 
the compactified radial coordinate $x$,
\begin{equation}
    x = \frac{r}{1+r} \, .
\label{comp_coord}
\end{equation}

The ADM mass of the configuration $M$ is given by 
\be
\label{expres_mass}
M=\frac{1}{2}\lim_{r\to\infty}r^2\partial_r N = \frac{1}{2}\lim_{x\to 1}\partial_x N ,
\ee
where the second expression gives the mass in terms of the
compactified coordinate $x$ from Eq.~\re{comp_coord}.
Alternatively, the mass of the system can also be found from the $(^0_0)$-component
of the stress-energy tensor \re{SET_tot},
\be
M=\alpha^2 \int_0^\infty T_0^0 r^2 dr .
\nonumber
\ee
This latter expression is employed to monitor the accuracy
of the numerical calculations.

Technically, the equations \re{GR1} - \re{Dirac_eq2} are discretized on a grid
consisting usually of about 400 grid points, 
but in some cases even 1000 or more grid points have been used.
The emerging system of nonlinear algebraic equations 
is then solved using a modified Newton method.
The underlying linear system is solved 
with the Intel MKL PARDISO sparse direct solver \cite{pardiso} 
and the CESDSOL library\footnote{Complex Equations-Simple Domain 
partial differential equations SOLver, a C++ package developed by I.~Perapechka,
see Refs.~\cite{Kunz:2019sgn,Herdeiro:2019mbz,Herdeiro:2021jgc}.}.
The package provides an iterative procedure for obtaining an exact solution 
starting from an initial guess configuration,
which we take to be the self-gravitating Skyrmion in the Yukawa decoupled limit 
(see, e.g., 
Refs.~\cite{Glendenning:1988qy,Bizon:1992gb,Heusler:1991xx,Heusler:1993ci}).

\subsection{Numerical results}

Let us start the discussion of the numerical results
by recalling the dependence of the gravitating Skyrmions on the 
effective gravitational coupling $\alpha^2= 4 \pi G f_\pi^2$,
i.e., when we set the Yukawa coupling to zero $(h = 0)$.
In that case, one obtains two branches of solutions
for the purely bosonic configurations
\cite{Glendenning:1988qy,Bizon:1992gb,Heusler:1991xx,Heusler:1993ci}.
Starting from the flat space Skyrmion at $\alpha=0$, the first branch
extends up to a maximal value of the effective gravitational coupling, $\alpha_{\text{max}}^2 \approx 0.0404$.
We refer to this branch as the Skyrmion branch.

At $\alpha_{\text{max}}$ this branch bifurcates with a second
branch, that is higher in mass and reaches all the way back 
to $\alpha=0$.
In fact, this limit is approached as $f_\pi \to 0$.
In the limit, the regular Einstein-Yang-Mills BM solution \cite{Bartnik:1988am}
is reached, as seen after appropriate rescaling~\cite{Bizon:1992gb}.
Thus we refer to this second branch as the BM branch.
In terms of the scaled ADM mass $M/\alpha^2$
one observes a decrease along both branches, 
with the minimal value of the scaled mass
being approached at $\alpha_{\text{max}}$.

\subsubsection{Spectral flow}

\begin{figure}[t]
    \begin{center}
        \includegraphics[width=.49\linewidth]{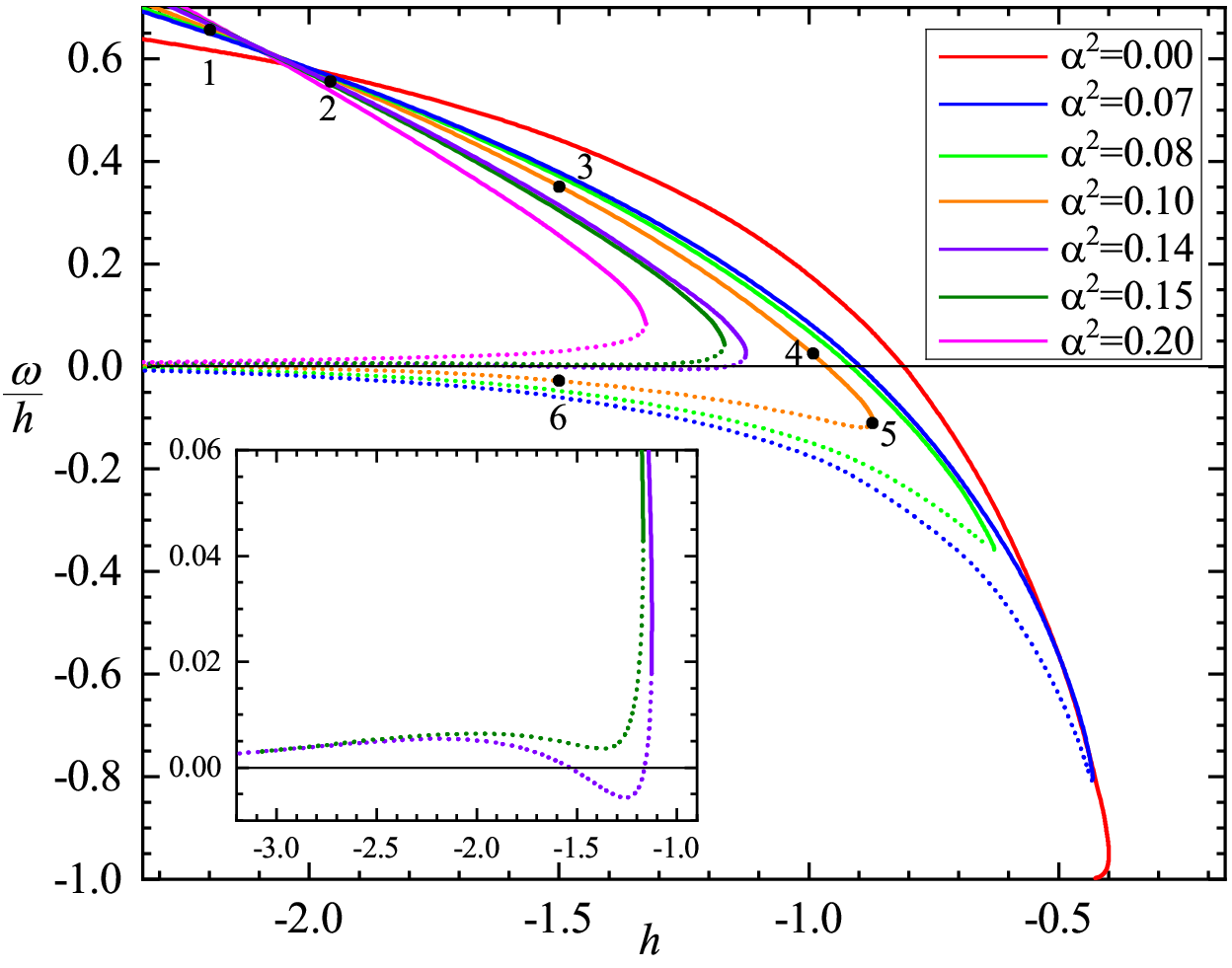}
        \includegraphics[width=.49\linewidth]{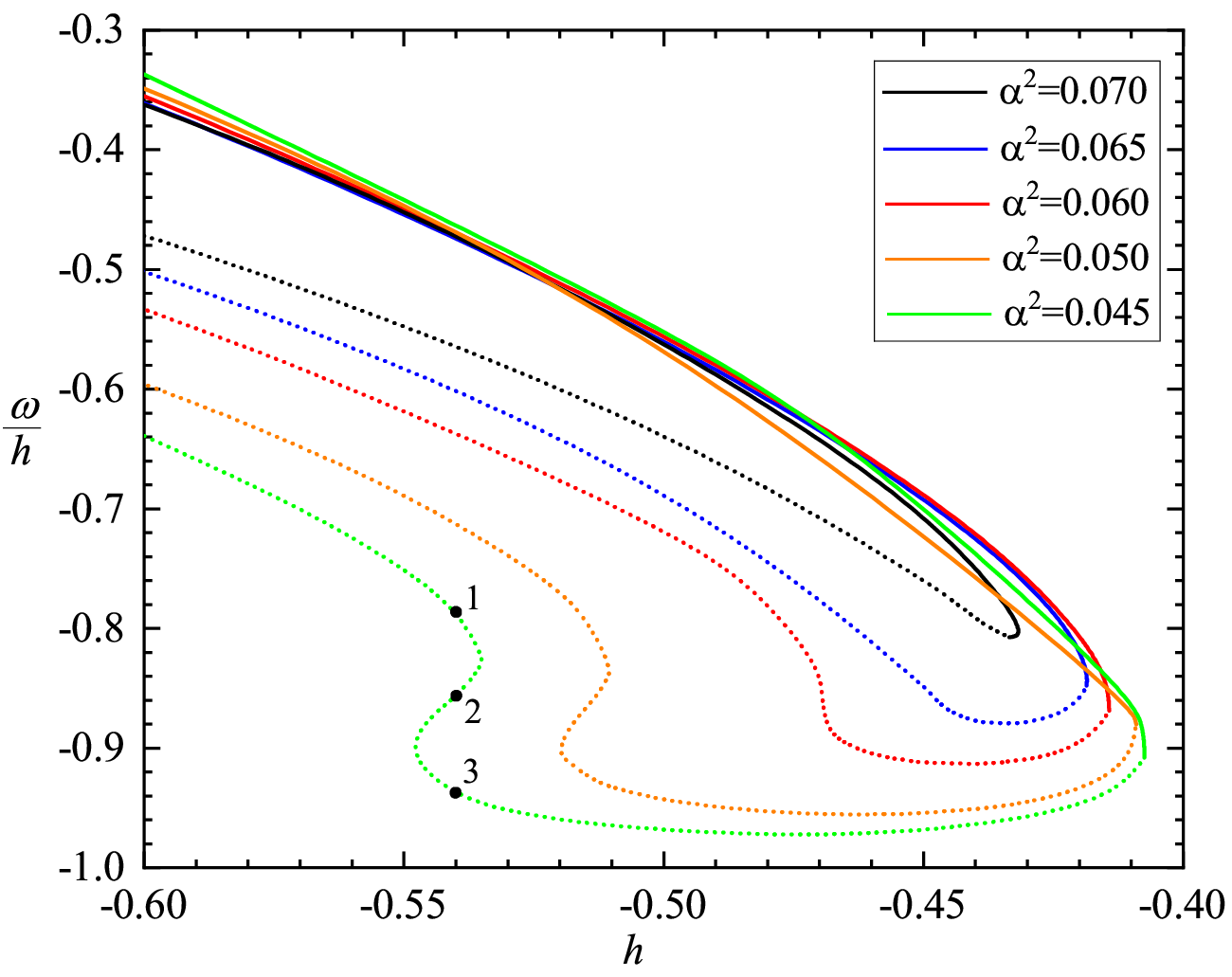}
        \vspace{-1.cm}
    \end{center}
    \caption{
        The scaled eigenvalue $\omega/h$ of the localized fermionic mode
        is shown versus the Yukawa coupling $h$ 
        for several values of the effective coupling $\alpha^2$.
        The solid lines represent the Skyrmion branches, while the dotted
        lines correspond to the BM and deformed BM branches.
        Note that the positive continuum resides at $\omega/h=-1$,
        since $h<0$.
        a) 
        The change of the spectral flow with increasing $\alpha^2$ is
        illustrated.
        The set of points 1 - 6 along the $\alpha^2=0.1$ branches represent
        illustrative configurations, whose matter and metric profile functions 
        are shown in Fig.~\ref{fig_field_distr}.
        The inset highlights the presence of multiple branches
        close to the transition of the bifurcation point from
        positive to negative eigenvalues $\omega$.
        b) 
        The multibranch structure for small values of $\alpha$ is illustrated.
        The set of points 1 - 3 along the $\alpha^2=0.045$ branches 
        (corresponding to Yukawa coupling $h=-0.54$) represents configurations,
        whose matter and metric profile functions are shown in 
        Fig.~\ref{fig_field_distr_2}.
        }
    \label{fig_freq_coupling}
\end{figure}

Turning next to the presence of the fermions,
we recall that in the limit of effective coupling $\alpha=0$
(taken in the limit of gravitational coupling $G=0$),
the Skyrmion-fermion system features spectral flow in Minkowski spacetime
\cite{Hiller:1986ry,Kahana:1984be,Balachandran:1998zq}.
This is illustrated by the red solid curve in Fig.~\ref{fig_freq_coupling}a,
where the fermion eigenvalue $\omega$, scaled by the Yukawa coupling $h$,
is shown versus $h$ for several values of $\alpha$,
starting from the Minkowski limit $\alpha=0$.
In that case, the fermionic mode emerges from the positive continuum
at a Yukawa coupling of $h_{\text{cr}}\approx -0.40$.
Note that the positive continuum resides in the figure at 
$\omega/h=-1$, since the Yukawa coupling $h$ is negative.

As the Yukawa coupling decreases below $h_{\text{cr}}$, the eigenvalue $\omega$ decreases, i.e., the scaled eigenvalue $\omega/h$ increases.
The eigenvalue $\omega$ then reaches zero at some particular value
of the Yukawa coupling $h$.
With the further decrease of $h$ the eigenvalue $\omega$ decreases further 
toward the negative continuum.
Thus, we observe a single fermionic mode
flowing monotonically from the positive 
to the negative continuum, in agreement with the index theorem. 

In the absence of gravity, the only parameter affecting the profile function 
of the Skyrmion due to the backreacting fermion is the Yukawa coupling.
The profile function always decreases monotonically 
from $F(0)=\pi$ to $F(\infty)=0$.
However, the effective size of the configuration and the mass
decrease as the magnitude of the coupling becomes stronger.
While there are also other bound modes in the spectrum of the Dirac fermions 
localized on the Skyrmion in Minkowski spacetime 
\cite{Hiller:1986ry,Kahana:1984be,Balachandran:1998zq}, 
the normalizable bound mode crossing zero is unique.

Let us next include gravity
and consider the dependence of the eigenvalue $\omega$ on
the Yukawa coupling $h$ for a set of values of 
the effective coupling $\alpha$.
One expects that instead of a single branch of 
Skyrmion-fermion solutions now
two branches of solutions should arise.
This expectation is due to the fact that for finite $\alpha$ 
the fermion may be localized and backreacting
on the solutions of either the Skyrmion branch or of the BM branch.

As seen in Fig.~\ref{fig_freq_coupling},
the presence of gravity indeed changes the picture 
accordingly and thus dramatically.
The single zero crossing fermionic level present in Minkowski spacetime,
that is related to the topology of the Skyrmion,
thus undergoes a significant change as the Skyrmion-fermion
system evolves toward the BM solution.

As the effective coupling $\alpha$ increases from zero, 
there arise indeed two branches of solutions,
that we will refer to as well as Skyrmion and BM branches.
We display these branches with different line styles 
in Fig.~\ref{fig_freq_coupling} and subsequent figures,
showing the Skrymion branches with solid curves and the
BM branches with dotted curves.

Now the fermionic mode no longer emerges from the positive continuum.
But for each $\alpha$ there is a maximal value of the Yukawa coupling
$h_{\text{max}}(\alpha) < h_{\text{cr}}$,
where the branches of solutions arise.
The fermionic eigenvalue $\omega$ at the bifurcation point of the branches 
decreases with decreasing $h$, passing zero at some point.
Although the bifurcation point then resides at $\omega<0$ 
for larger values of $\alpha$, the eigenvalues still approach
zero with decreasing $h$ along the BM branch.

In fact, the inset in Fig.~\ref{fig_freq_coupling}a shows
that in a small parameter range, the observed pattern is slightly more
complicated.
We do not only see the Skyrmion branch, obtained by varying $G$,
and the BM branch, obtained by varying $f_\pi$, while
$\alpha$ is the same on both branches.
But there arises a small intermediate branch in the vicinity
of the transition of the eigenvalue $\omega$ to purely negative values.
This branch is also displayed with dotted line style,
and in combination with the BM branch
this will be referred to as a deformed BM branch.

We associate the reason for this subtlety with our considerations above,
where we introduced the dimensionless Yukawa coupling $h\to h/(a_0 f_\pi)$. 
Hence, the same value of the coupling $h$ may correspond to 
different choices of all three parameters, 
and therefore one may expect multiple branches of solutions to appear. 
Since we fix the normalization condition in our numerical analysis 
to a particular value of the Skyrme parameter $a_0$, 
the branches of solutions may be related to the variation 
of the two remaining parameters, $h$ and $f_\pi$. 
In turn, variations of the latter parameter may be related to
a corresponding change of the effective gravitational constant $\alpha$.

When we consider very small values of $\alpha$, 
where the eigenvalues of the Dirac operator 
get close to the positive continuum,
this phenomenon of multiple branches is seen again,
and is illustrated in Fig.~\ref{fig_freq_coupling}b.
Here a more detailed analysis of the critical behavior of the solutions 
reveals these interesting features, as $\alpha^2 < 0.06$.
Then there is no longer a single bifurcation point, 
at which the Skyrmion and BM branches merge,
but again multiple branches appear,
deforming the BM branch.
For example, for $\alpha^2=0.045$ there are three bifurcation points, 
as seen in Fig.~\ref{fig_freq_coupling}b.
The figure also shows that as $\alpha^2$ is increased above $0.06$ the multibranch pattern with several bifurcation points disappears
and is replaced by a single bifurcation point with double branch structure.

\subsubsection{Field configurations}

\begin{figure}[t]
    \begin{center}
        \includegraphics[width=1.\linewidth]{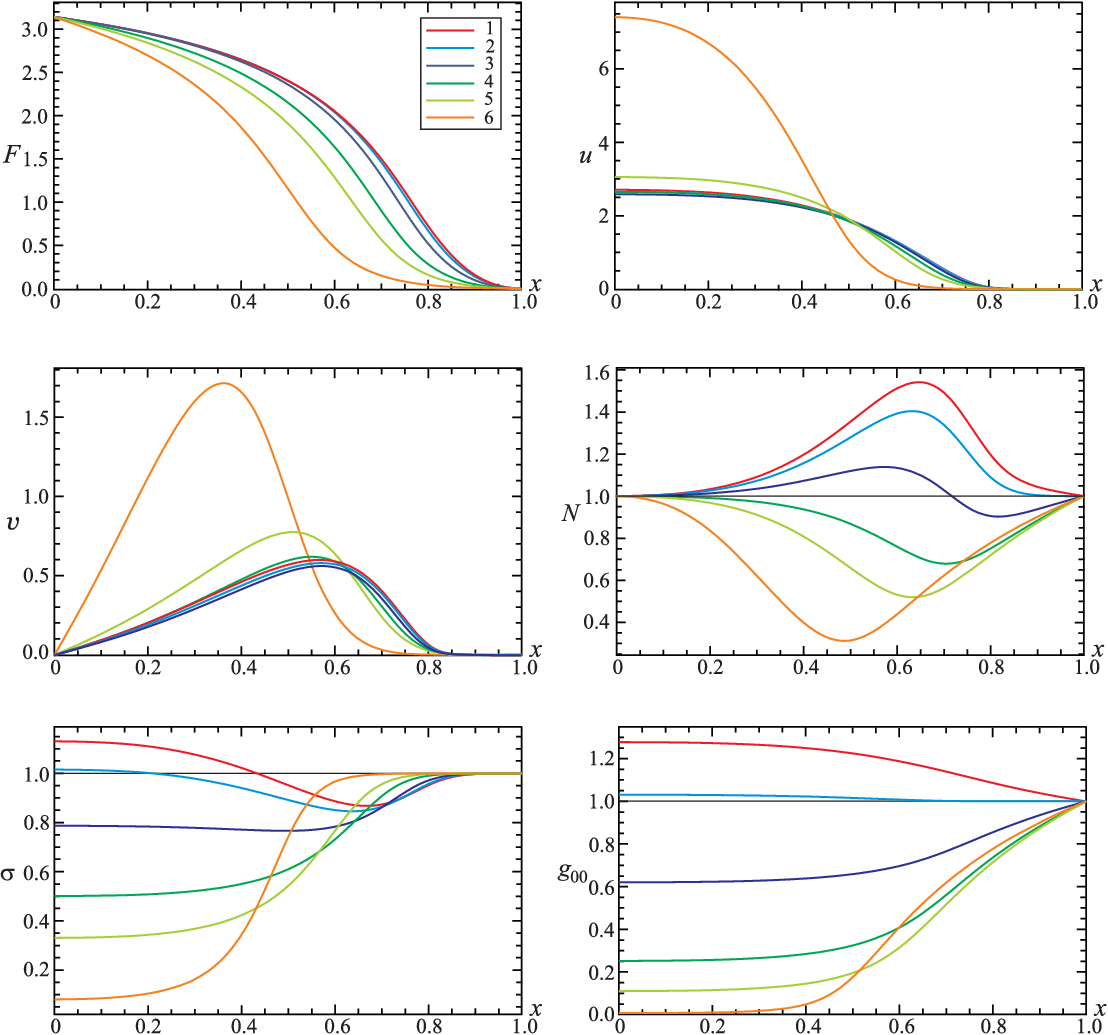}
        \vspace{-1.cm}
    \end{center}
    \caption{The profile functions $F$, $u$, $v$, $N$, and $\sigma$ of the gravitating Skyrmion-fermion system 
    are shown versus the compactified radial coordinate $x$
    for $\alpha^2=0.1$ for the points 1 - 6 in Fig.~\ref{fig_freq_coupling}a.
    The solution labeled by point 2 corresponds to zero ADM mass.
    }
    \label{fig_field_distr}
\end{figure}

\begin{figure}[t]
    \begin{center}
        \includegraphics[width=1.\linewidth]{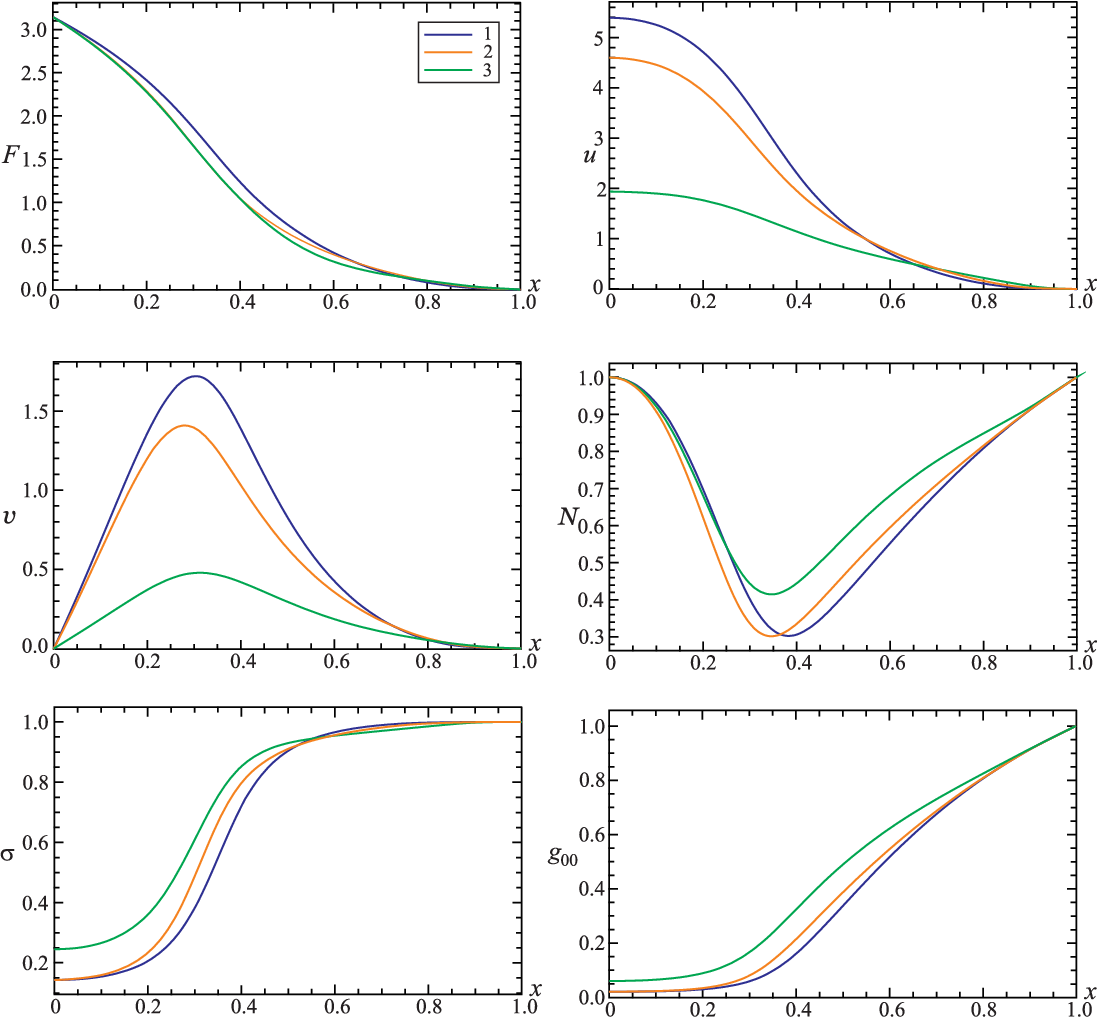}
        \vspace{-1.cm}
    \end{center}
    \caption{The profile functions $F$, $u$, $v$, $N$, and $\sigma$ of the gravitating Skyrmion-fermion system 
    are shown versus the compactified radial coordinate $x$
    for $\alpha^2=0.045$ for the points 1 - 3 in Fig.~\ref{fig_freq_coupling}b.
        }
    \label{fig_field_distr_2}
\end{figure}

In order to get a better understanding of the backreacting Skrymion-fermion systems
and the associated spectral flow, let us now inspect some field configurations.
We first consider the configurations labeled by points 1 - 6
in Fig.~\ref{fig_freq_coupling}a.
These points correspond to configurations with $\alpha^2=0.1$ and several values
of the Yukawa coupling $h$.
Points 1 - 4 reside on the Skyrmion branch, point 5 represents the
configuration at the bifurcation of the branches, and point 6
resides on the BM branch.

We exhibit the respective sets of 
profile functions $F$, $u$, $v$, $N$, and $\sigma$ together
with the metric component $g_{00}$ in Fig.~\ref{fig_field_distr}.
The configuration on the BM branch (point 6) possesses the smallest size.
This is not unexpected, since the BM solution itself would shrink 
to a point in the present coordinate system.
The fermion functions $u$ and $v$ then need to reach large values in the
inner region to satisfy the normalization condition.
Toward the bifurcation of the branches (point 5) the size of the configurations
increases, and then continues to increase along the Skyrmion branch.

We further observe that as the fermion eigenvalue $\omega$ crosses zero (point 4) 
and then becomes negative,
the metric functions $N(r)$ and $\sigma(r)$ possess increasing
the minimal values.
Interestingly, when $h$ is further decreased along the Skyrmion branch,
the metric function $N(r)$ is no longer bounded by its asymptotic values
from above.
Instead, $N(r)$ increases above unity, first only in the inner part (point 3),
and then everywhere in the interior (points 2 and 1),
as seen in Fig.~\ref{fig_field_distr}.
Clearly, this has effects on the mass of the configurations,
as discussed below.

We display in Fig.~\ref{fig_field_distr_2} the profile functions of the fields 
of the backreacting Skyrmion-fermion system for
configurations in the multibranch region for small $\alpha$,
close to the positive continuum.
In particular, we show the three sets of functions obtained 
for $\alpha^2=0.045$ and $h=-0.54$, that are labeled as points 1 - 3 in Fig.~\ref{fig_freq_coupling}b.
The figure clearly shows that the solutions become increasingly localized,
as we move to smaller fermion eigenvalues along the
deformed BM branch.
On the other hand, the ADM mass of these three solutions remains almost the same.

\subsubsection{ADM mass}

\begin{figure}[t]
\begin{center}
\mbox{ 
    \includegraphics[width=.5\linewidth]{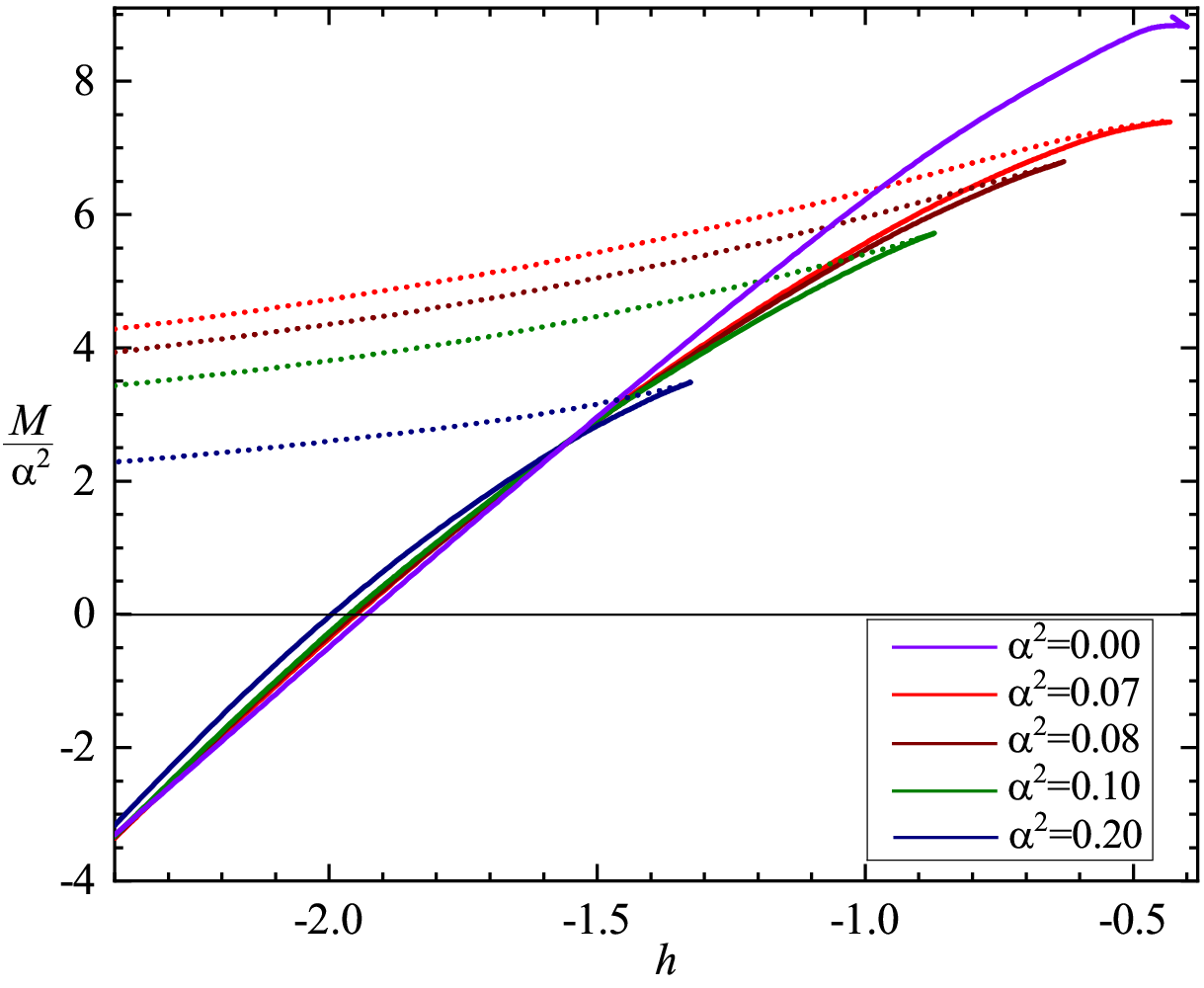}
    \includegraphics[width=.48\linewidth]{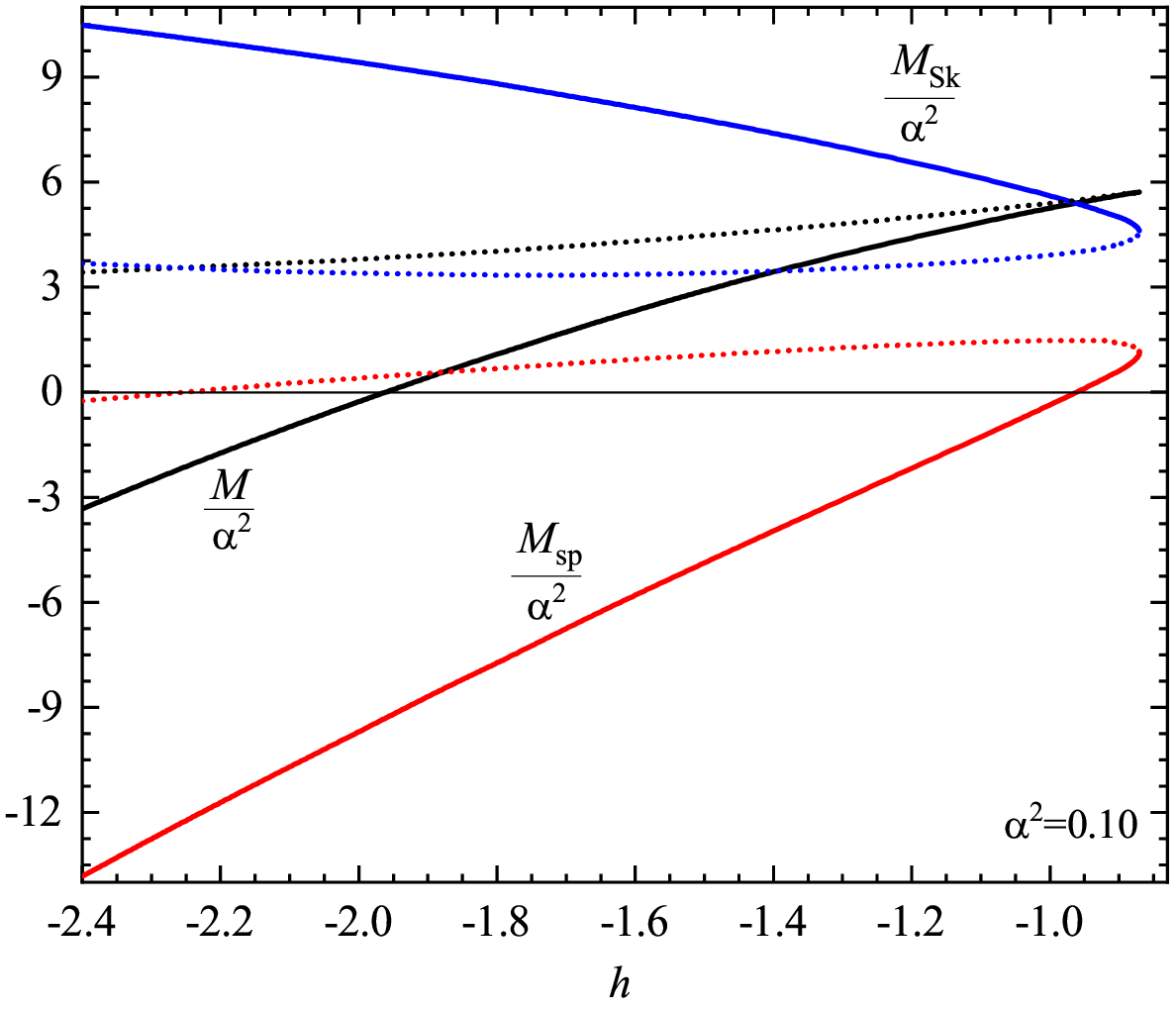}
    }
\vspace{-1.cm}
\end{center}
    \caption{
    a) The scaled ADM mass $M/\alpha^2$ of the gravitating Skyrmion-fermion system
    is shown versus the Yukawa coupling $h$ for a set of values 
    of the effective coupling $\alpha^2$.
    b) The contributions to the scaled total ADM mass ($M/\alpha^2$, black) 
    from the fermion ($M_{\text{sp}}/\alpha^2$, red) 
    and the Skyrme ($M_{\text{Sk}}/\alpha^2$, blue) fields 
    are shown versus the Yukawa coupling $h$ for effective coupling
    $\alpha^2=0.1$. 
}
    \label{fig_Mass_contrib}
\end{figure}

We exhibit the ADM mass of the Skyrmion-fermion system in 
Fig.~\ref{fig_Mass_contrib}a,
where we display the scaled mass $M/\alpha^2$ 
versus the Yukawa coupling $h$ for a set of values
of the effective coupling $\alpha$, also shown in Fig.~\ref{fig_freq_coupling}.
The figure highlights again the bifurcations of the Skyrmion branches (solid)
with the BM branches (dotted).
As noted above, the maximal value of the Yukawa coupling is reached in the flat
spacetime limit ($h\approx -0.4$), and decreases with increasing 
$\alpha$.

For a given value of the Yukawa coupling $h$, 
the ADM mass of the configurations on the Skyrmion branch is smaller
than the mass on the BM branch.
With increasing $\alpha$ the mass of the configurations on the BM branches, 
including the bifurcation points, decreases.
In contrast, the mass on the lower parts of the Skyrmion branches
does not vary significantly as $\alpha$ is increased.

Intriguingly, the ADM mass on the Skyrmion branches
crosses zero when the Yukawa coupling is decreased, 
as demonstrated in Fig.~\ref{fig_Mass_contrib}.
Inspection of the configurations at this critical point with $M=0$
shows that the metric component $g_{00}$ is nearly unity
almost everywhere in space,
and that the first derivative of the metric function $N$ 
at spatial infinity vanishes.
This is illustrated in Fig.~\ref{fig_field_distr} by the functions 
corresponding to point 2.

Beyond this critical point, the ADM mass 
of the backreacting Skyrmion-fermion system becomes \emph{negative}
as the Yukawa coupling $h$ decreases along the Skyrmion branch,
as seen in Fig.~\ref{fig_Mass_contrib}.
While surprising at first, this hints at the capacity of fermions
to violate the energy conditions.
In contrast to this intriguing behavior of the ADM mass on the Skyrmion branch, 
the mass remains always positive along the BM branch.

To get further insight, let us compare the contributions of the 
Skyrmion and the fermion fields to the total mass.
We present in Fig.~\ref{fig_Mass_contrib}b the dependence 
of the corresponding scaled quantities 
$M_{\text{Sk}}/\alpha^2$ and $M_{\text{sp}}/\alpha^2$  
on the Yukawa coupling $h$ for $\alpha^2=0.10$,
together with the scaled total mass 
$M/\alpha^2=(M_{\text{Sk}}+M_{\text{sp}})/\alpha^2$.
These mass contributions are obtained by evaluating 
$M_{\text{Sk}} = \alpha^2\int dr\, r^2 (T_0^0)_\text{Sk}$ 
and $M_{\text{sp}} = \alpha^2\int dr\, r^2 (T_0^0)_\text{sp}$
(with $(T_0^0)_\text{Sk}$ and $(T_0^0)_\text{sp}$ 
given by Eqs.~\re{T00} and \re{T00_sp}, respectively).
Clearly, the negative contribution of the fermionic mode becomes dominant 
on the Skyrmion branch, rendering the total mass negative.
On the BM branch, on the other hand, both contributions remain positive 
(for $h > -2.25$). 

When we consider the scaled ADM mass $M/\alpha^2$ for configurations
obtained by varying the effective coupling $\alpha$ 
for a set of fixed values of the Yukawa coupling $h$,
we again observe two branches of solutions, a Skyrmion branch
and a BM branch, that bifurcate at some maximal value $\alpha_{\text{max}}(h)$.
As the Yukawa coupling $h$ decreases, $\alpha_{\text{max}}(h)$ increases.
Curiously, at a particular value of the Yukawa coupling ($h=-1.565$)
the scaled mass $M/\alpha^2$ remains approximately constant along the 
Skyrmion branch \cite{Dzhunushaliev:2024iag}.

\subsubsection{Violation of the null and weak energy conditions}

We now demonstrate that the intriguing behavior of the solutions 
of the system \re{GR1}~-~\re{Dirac_eq2} on the Skyrmion branches 
is related to the violation of the null and weak energy conditions.
These demand that the stress-energy tensor $T_{\mu\nu}$ 
of the system satisfies the following inequalities,
$$
    T_{\mu\nu} k^\mu k^\nu \geq 0 \quad \text{and}\quad T_{\mu\nu} V^\mu V^\nu \geq 0 ,
$$
for any light-like vector $k^\mu$, $g_{\mu\nu}k^\mu k^\nu =0$, 
and for any timelike vector $V^\mu$, $g_{\mu\nu}V^\mu V^\nu >0$, 
respectively (for a review, see, e.g., Ref.~\cite{Rubakov:2014jja}).
For the system \re{Lag-tot} and
the spherically symmetric Ansatz \re{metrics} - \re{spherical-fermions},
the only non-vanishing components of the stress-energy tensor~\re{T-Sk}
of the Skyrme field are
\begin{align}
    \left( T^0_0\right)_{\text{Sk}} = & N \frac{
        2 \sin^2F + r^2 }{2r^2} {F'}^2
        + \frac{\sin^2 F + 2 r^2 }{2 r^4} \sin^2F ,
\label{T00}\\
    \left( T^1_1\right)_{\text{Sk}} = & - N \frac{
    2 \sin^2 F + r^2 }{2 r^2} {F'}^2
    + \frac{\sin ^2 F + 2 r^2 }{2 r^4} \sin^2 F ,
\label{T11}\\
    \left( T^2_2\right)_{\text{Sk}} = & \left( T^3_3\right)_{\text{Sk}} = \frac{1}{2} N {F'}^2
    - \frac{\sin^4 F}{2 r^4} ,
\label{T22}
\end{align}
and of the stress-energy tensor \re{T-sp} of the fermion field are
\begin{align}
    \left( T^0_0\right)_{\text{sp}} = & \omega \frac{ u^2 + v^2}{\sqrt{N} \sigma } ,
\label{T00_sp}\\
    \left( T^1_1\right)_{\text{sp}} = & 2 u v  \frac{h r \sin F + 1}{r}
    + h \cos F \left(v^2 - u^2\right)
    - \omega \frac{u^2 + v^2}{\sqrt{N} \sigma} ,
\label{T11_sp}\\
    \left( T^2_2\right)_{\text{sp}} = & \left( T^3_3\right)_{\text{sp}} = -\frac{u v}{r} .
\label{T22_sp}
\end{align}
These components are given in terms of dimensionless quantities 
(following the rescaling $T^\mu_\nu \to T^\mu_\nu/(a_0^2 f_\pi^4)$).
The null and the weak energy conditions for the 
backreacting Skyrmion-fermion system then become
\begin{equation}
\begin{split}
    \epsilon + p \equiv & T^0_0 - T^1_1 =
    N \frac{ 2 \sin ^2 F + r^2}{r^2} {F'}^2
\\
    &
    + h \left[
        \cos F \left(u^2 - v^2\right) - 2 u v \sin F
    \right]
    + 2 \omega \frac{u^2 + v^2}{\sqrt{N} \sigma}
    - \frac{2 u v}{r} \geq 0,
\label{NEC}
\end{split}
\end{equation}
where $\epsilon$ denotes the energy density and $p$ the radial pressure.
The weak energy condition also implies $\epsilon\equiv T_0^0\geq 0$.

\begin{figure}[t]
\begin{center}
    \includegraphics[width=1.\linewidth]{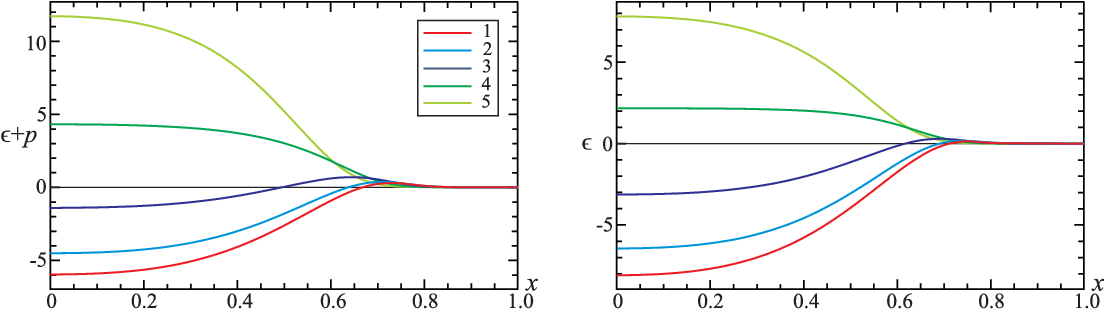}
\vspace{-1.cm}
\end{center}
    \caption{The validity, respectively violation, 
    of the null and weak energy conditions 
    is demonstrated for the Skyrmion-fermion solutions 
    labeled by points 1 - 5 in Fig.~\ref{fig_freq_coupling}a.
    a) The sum ($\epsilon + p$) of the energy density $\epsilon$ and the pressure $p$
    is shown versus the compactified radial coordinate $x$.
    b) The energy density $\epsilon$ is shown for the same set of solutions.}
    \label{fig_NEC}
\end{figure}

Evidently, the first term on the right-hand side of the expression \re{NEC} is related to the contribution of the Skyrme field.
It is always positive, and a violation of the null/weak energy conditions can only be related to the contribution of the Dirac field.

We display in Fig.~\ref{fig_NEC}a the combination $(\epsilon + p)$ 
on the right-hand side of the null/weak energy conditions~\re{NEC} 
and in Fig.~\ref{fig_NEC}b the energy density $\epsilon$ 
versus the compactified radial coordinate $x$ 
for a set of Skyrmion-fermion solutions.
In particular, we choose again the configurations considered earlier in detail,
that are labeled by the points 1 - 5 in Fig.~\ref{fig_freq_coupling}a
for $\alpha^2=0.1$.
Whereas the configurations at the bifurcation (point 5) and slightly further
on the Skyrmion branch (point 4) still respect the energy conditions,
these become violated as we progress along the Skyrmion branch
(points 3, 2, and 1). 
We also note that for all configurations 1~-~5 
the pressure $p$ is always positive 
and the violation of the null/weak energy conditions 
is only caused by the energy density.

\subsubsection{Localized fermions with finite bare mass}

\begin{figure}[t]
    \begin{minipage}[t]{.49\linewidth}
        \begin{center}
\includegraphics[width=1.\linewidth]{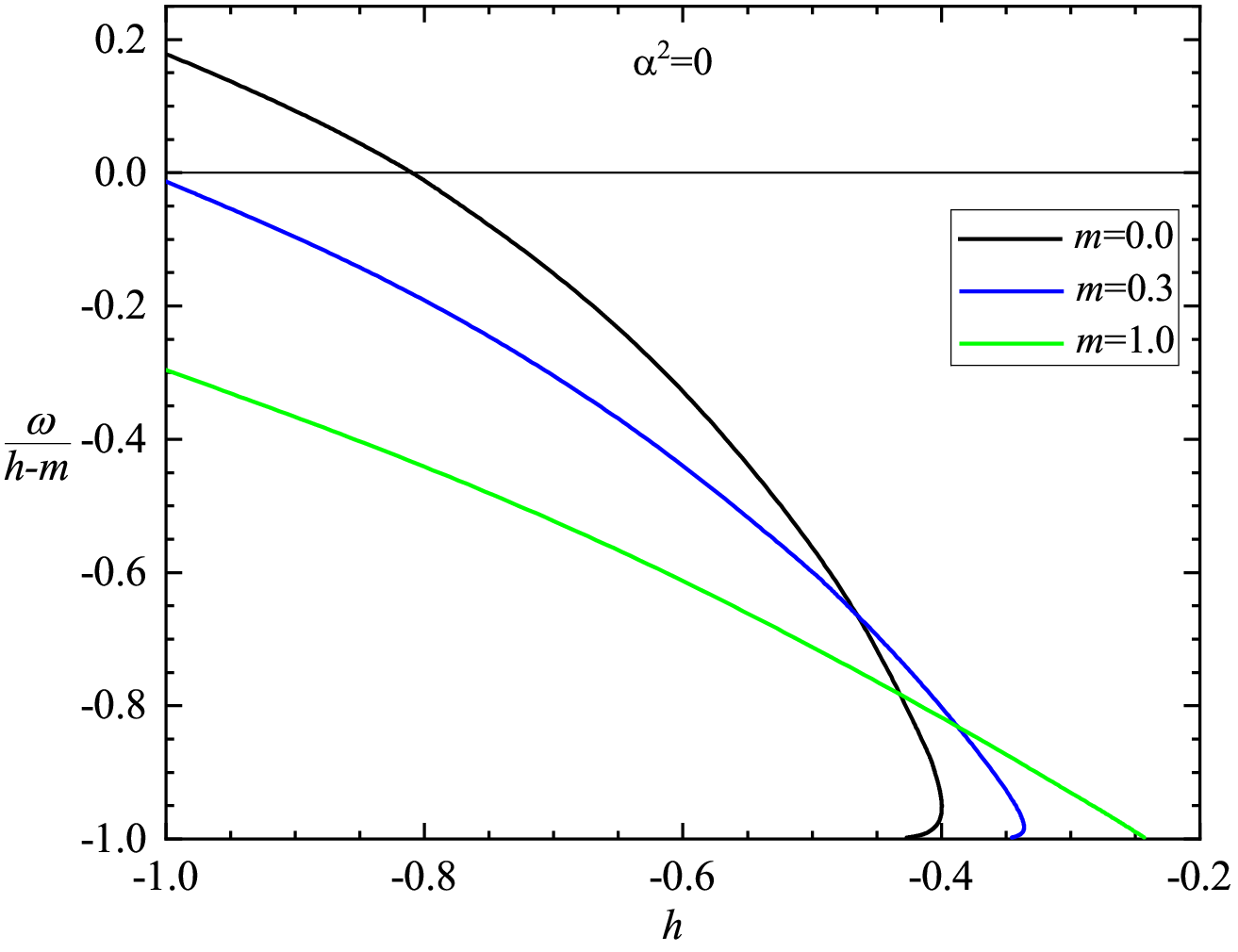}
        \end{center}
    \end{minipage}\hfill
    \begin{minipage}[t]{.49\linewidth}
        \begin{center}
\includegraphics[width=.98\linewidth]{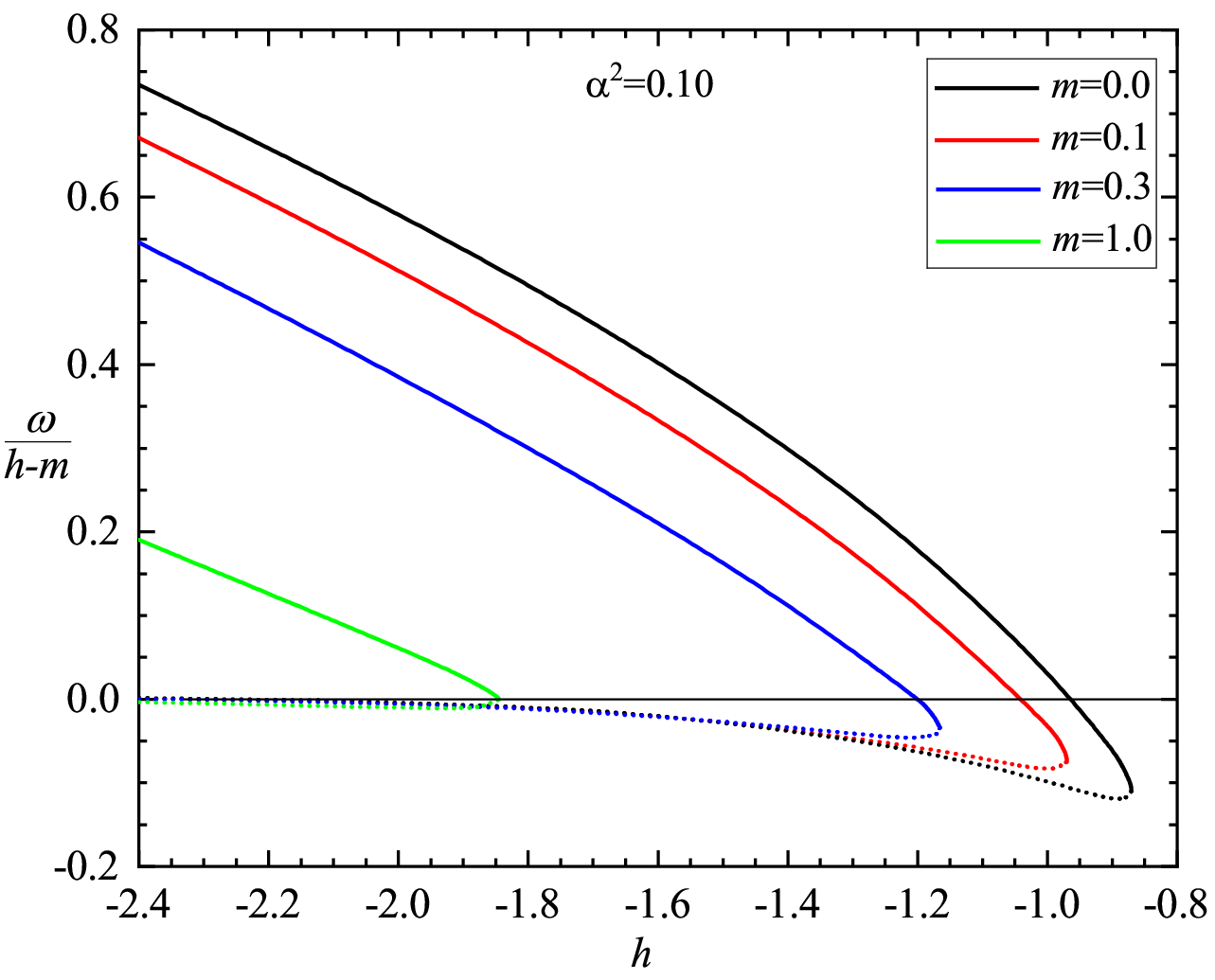}
\end{center}
    \end{minipage}
    \begin{minipage}[t]{.49\linewidth}
        \begin{center}
\includegraphics[width=1.\linewidth]{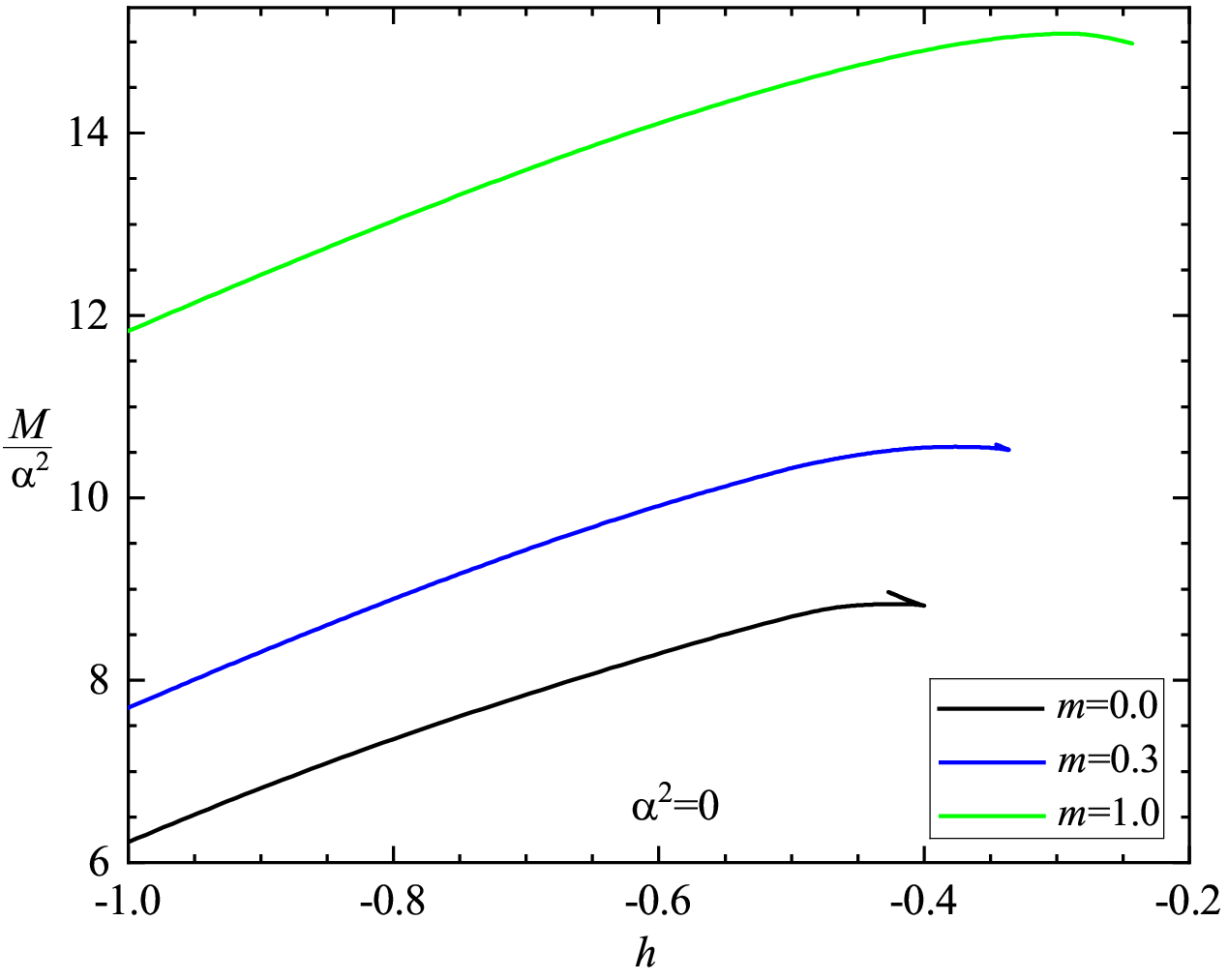}
        \end{center}
    \end{minipage}\hfill
    \begin{minipage}[t]{.49\linewidth}
        \begin{center}
\includegraphics[width=.99\linewidth]{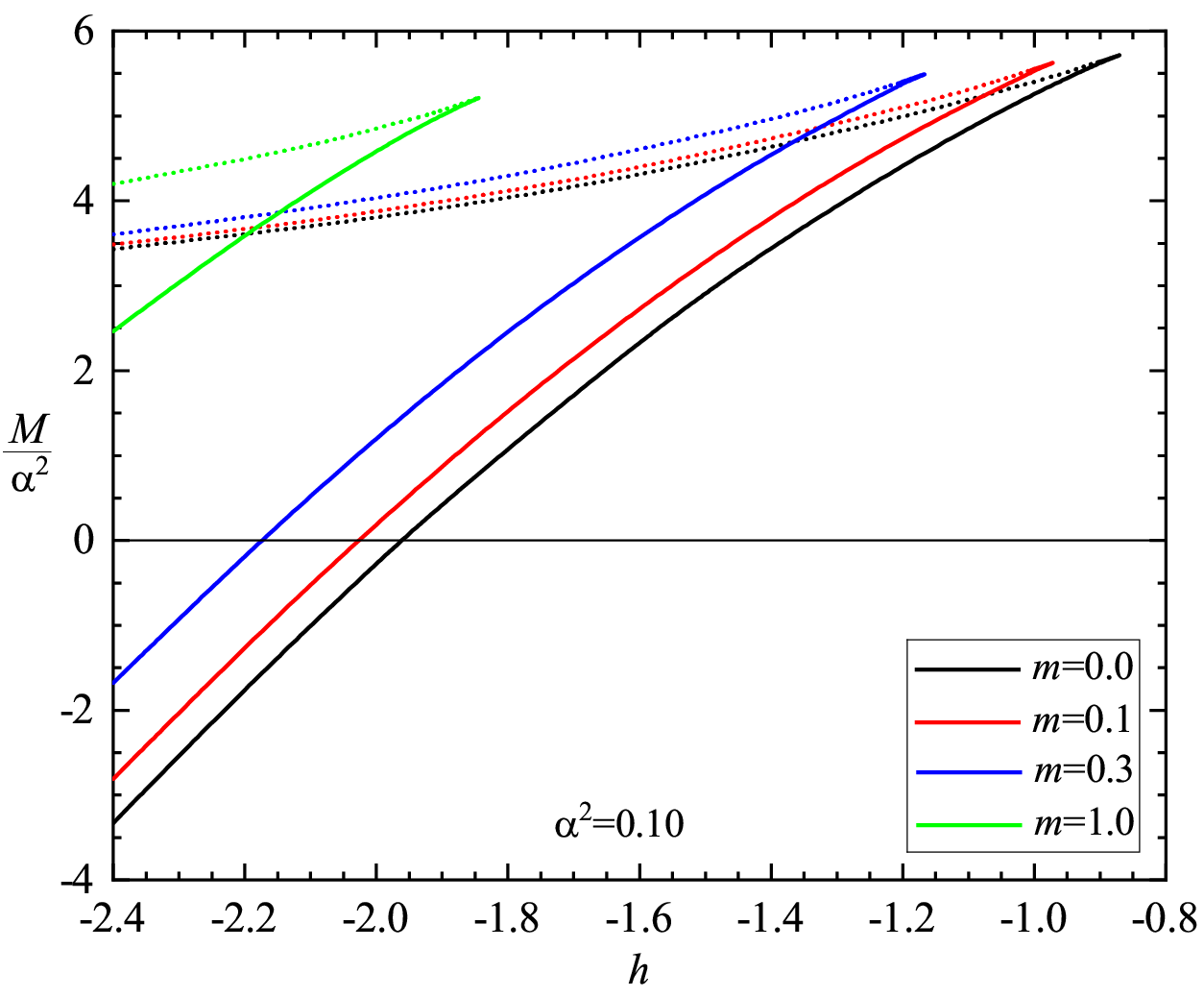}
\end{center}
    \end{minipage}
\caption{The influence of a finite bare mass $m$ of the fermions on 
        the backreacting Skyrmion-fermion system is demonstrated.
        a) The scaled eigenvalue $\omega/(h-m)$ of the localized fermionic mode
        is shown versus the Yukawa coupling $h$ for several values of
        the bare mass $m$ in Minkowski space ($\alpha^2=0$), and
        b) for effective coupling $\alpha^2=0.10$.
        c) The scaled ADM mass $M/\alpha^2$ is shown versus
        the Yukawa coupling $h$ for several values of
        the bare mass $m$ in Minkowski space ($\alpha^2=0$), and
        d)~for effective coupling $\alpha^2=0.10$.
        The solid lines represent the Skyrmion branches, while the dotted
        lines correspond to the BM branches.
}
\label{fig_with_bare_mass}
\end{figure}

We now address the influence of a finite value of the bare mass $m$ of the fermions
on the Skyrmion-fermion configurations with backreaction taken into account.
In particular, we consider several values of the bare mass $m$ for the system
in Minkowski space and in the presence of gravity,
and we exhibit our main findings in Fig.~\ref{fig_with_bare_mass}.

We note that the previously obtained pattern changes only slightly
when massive fermionic modes
are localized on the gravitating Skyrmion,
and the Yukawa coupling $h$ and the effective coupling $\alpha$ are varied.
First, we note that the asymptotic behavior of the fermion field 
is still given by Eq.~\eqref{asympt} up to the replacement
$h\to \bar{h}\equiv (h-m)$. 
The latter is necessary since the fermion obtains an effective mass 
$m_\text{eff}=m-h$ due to the Yukawa coupling.
Consequently, the emergence of the fermionic mode
from the positive continuum arises at $\omega/m_\text{eff} = 1$ 
or, as shown in Fig.~\ref{fig_with_bare_mass}a, 
at $\omega/(h-m) = - 1$.

Starting the discussion in Minkowski space ($\alpha^2=0$), 
we exhibit the scaled eigenvalue $\omega/(h-m)$ of the fermionic mode
versus the Yukawa coupling $h$ in Fig.~\ref{fig_with_bare_mass}a
for several values of the bare mass $m$, including
the vanishing bare mass case for comparison.
Clearly, the spectral flow exhibits the expected zero crossing behavior
of the eigenvalue also for massive fermions,
as the Yukawa coupling decreases.

However, as the bare mass increases from zero, 
also a new feature is observed. 
We note that for vanishing bare mass 
the scaled eigenvalue $\omega/h$ of the fermionic mode
emerges from the positive continuum slightly below $h_{\text{max}}$.
Thus the dependence of the scaled eigenvalue $\omega/h$
on the Yukawa coupling $h$ is not quite monotonic close to
the positive continuum.
With increasing bare mass this non-monotonic behavior 
of $\omega/(h-m)$ is reduced and then disappears, 
as seen in the figure for $m=0.3$ and $m=1.0$, respectively.
We also observe in Fig.~\ref{fig_with_bare_mass}a
that the critical value of the Yukawa coupling, 
where the localized mode emerges from the continuum, 
increases with increasing $m$.

In the presence of gravity
two branches of solutions arise
also for finite values of the fermion bare mass,
the Skyrmion branch and the BM branch.
This is illustrated in Fig.~\ref{fig_with_bare_mass}b
for effective coupling $\alpha^2=0.10$.
We observe that, with increasing $m$, 
the bifurcation point of the branches 
now arise at decreasing values of the Yukawa coupling $h$.
Thus solutions for the backreacting Skyrmion-fermion system 
exist only for decreasing $h_\text{max}(m)$.

Next, we exhibit the scaled mass $M/\alpha^2$ 
versus the Yukawa coupling $h$ for these sets of solutions 
in Fig.~\ref{fig_with_bare_mass}c for Minkowski space
and in Fig.~\ref{fig_with_bare_mass}d 
in the presence of gravity ($\alpha^2=0.10$).
We note that in Minkowski space
the scaled ADM mass of the configurations
increases significantly with increasing bare mass,
including the maximal value of the scaled ADM mass.

In contrast, when gravity is coupled to the system,
the maximal value of the ADM mass
depends only weakly on the bare mass $m$.
Overall, the ADM mass retains its characteristic behavior 
discussed above, with the branches shifting toward
smaller values of the Yukawa coupling $h$
with increasing bare mass.
Along the Skyrmion branch 
the mass always crosses zero and then turns negative.
Again, the energy conditions become violated 
along the Skyrmion branch for sufficiently small values of $h$.

\subsubsection{Radially excited fermionic modes}

\begin{figure}[t]
    \begin{minipage}[t]{.49\linewidth}
        \begin{center}
\includegraphics[width=1.\linewidth]{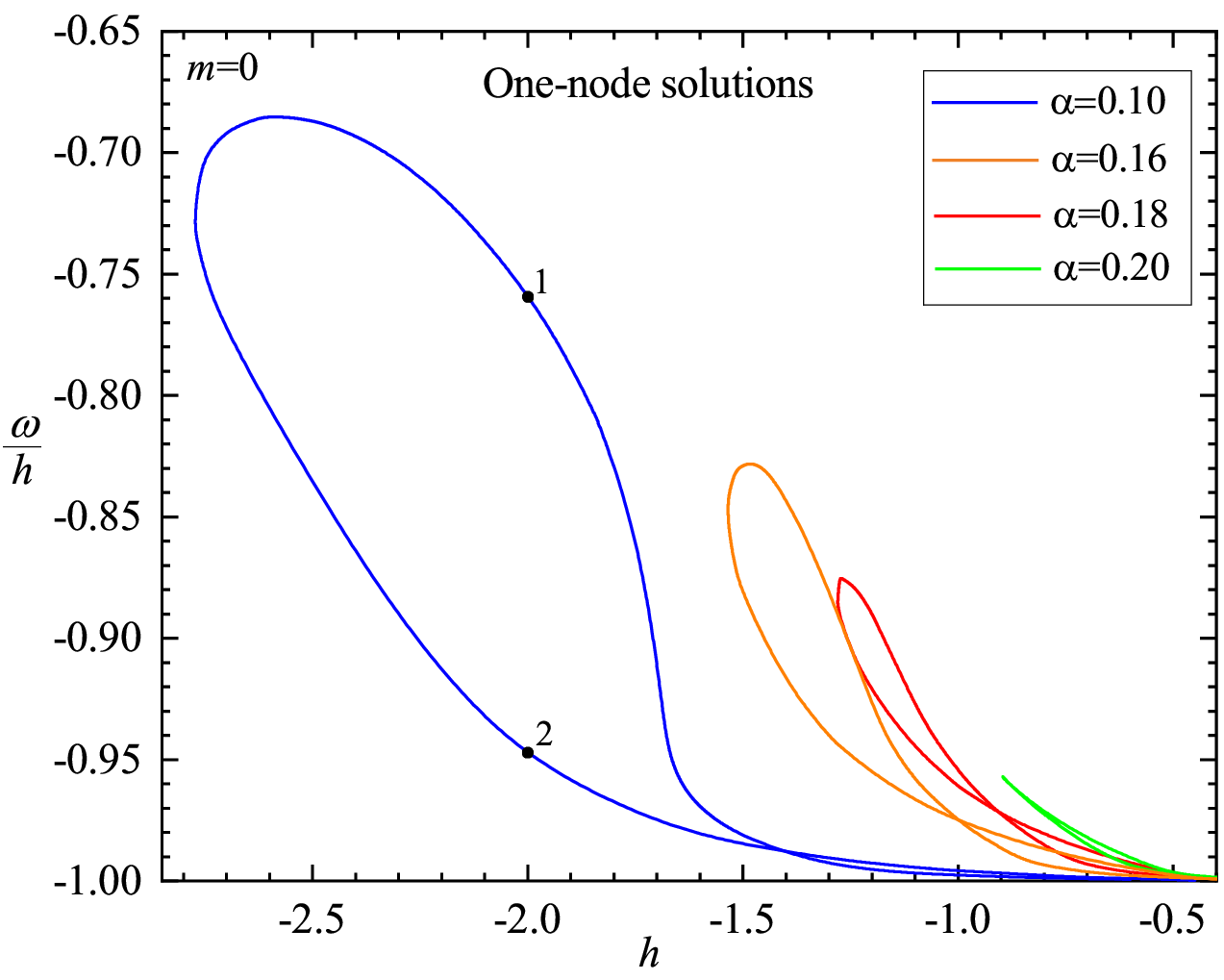}
        \end{center}
    \end{minipage}\hfill
    \begin{minipage}[t]{.49\linewidth}
        \begin{center}
\includegraphics[width=.98\linewidth]{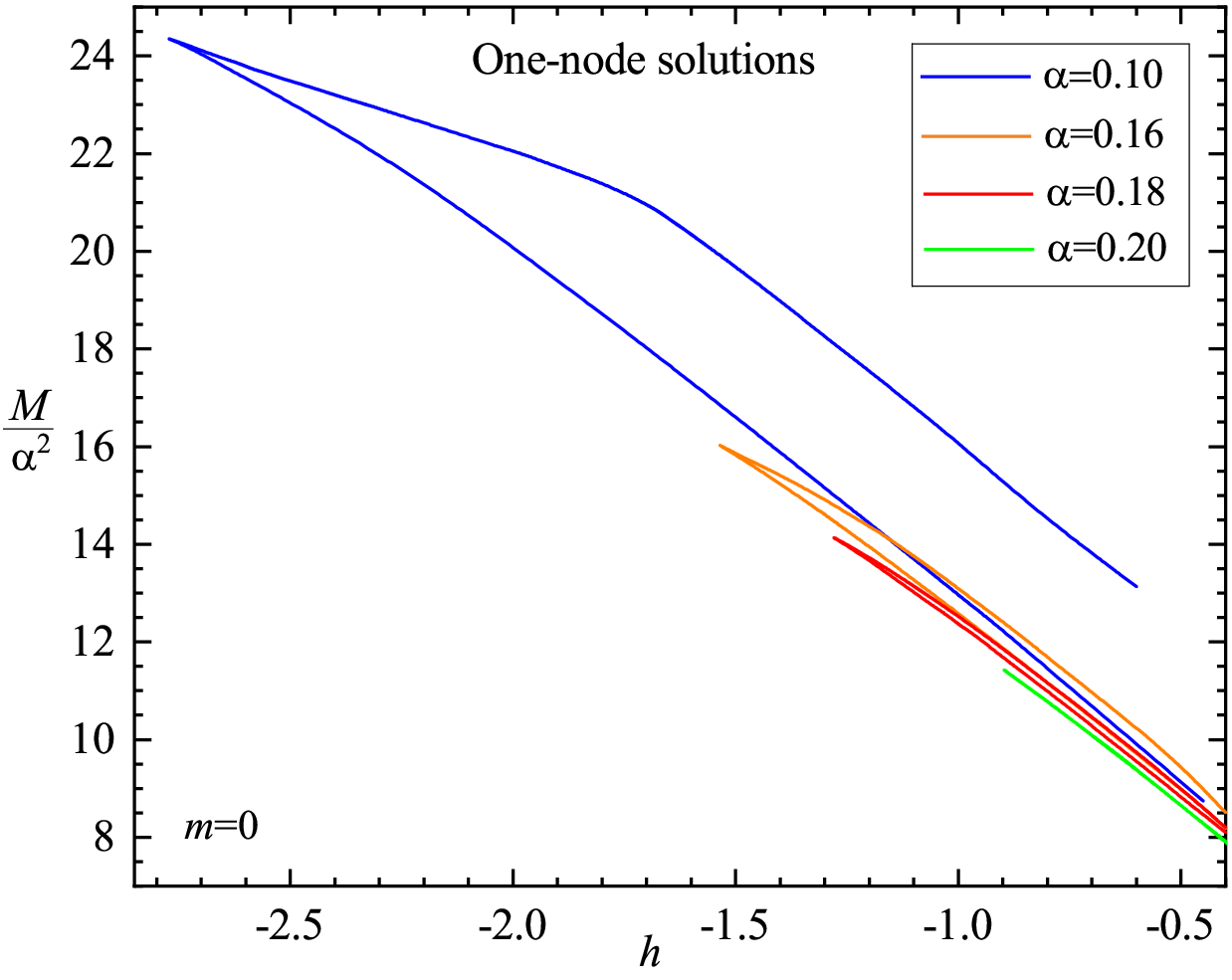}
\end{center}
    \end{minipage}
\caption{Spherically symmetric radially excited solutions are 
        exhibited for several values of the effective coupling $\alpha^2$.
        a) The scaled eigenvalue $\omega/h$ of the localized fermionic mode
        is shown versus the Yukawa coupling $h$. 
        Note that the positive continuum resides at $\omega/h=-1$.
        The points 1 and 2 on the $\alpha=0.1$ loop at $h=-2$ represent
        configurations, whose matter and metric profile functions 
        are shown in Fig.~\ref{fig_field_distr_3}.
        b) The scaled ADM mass $M/\alpha^2$ is shown 
        versus the Yukawa coupling $h$. 
}
\label{fig_freq_coupling_one_node}
\end{figure}

\begin{figure}[t]
    \begin{center}
        \includegraphics[width=1.\linewidth]{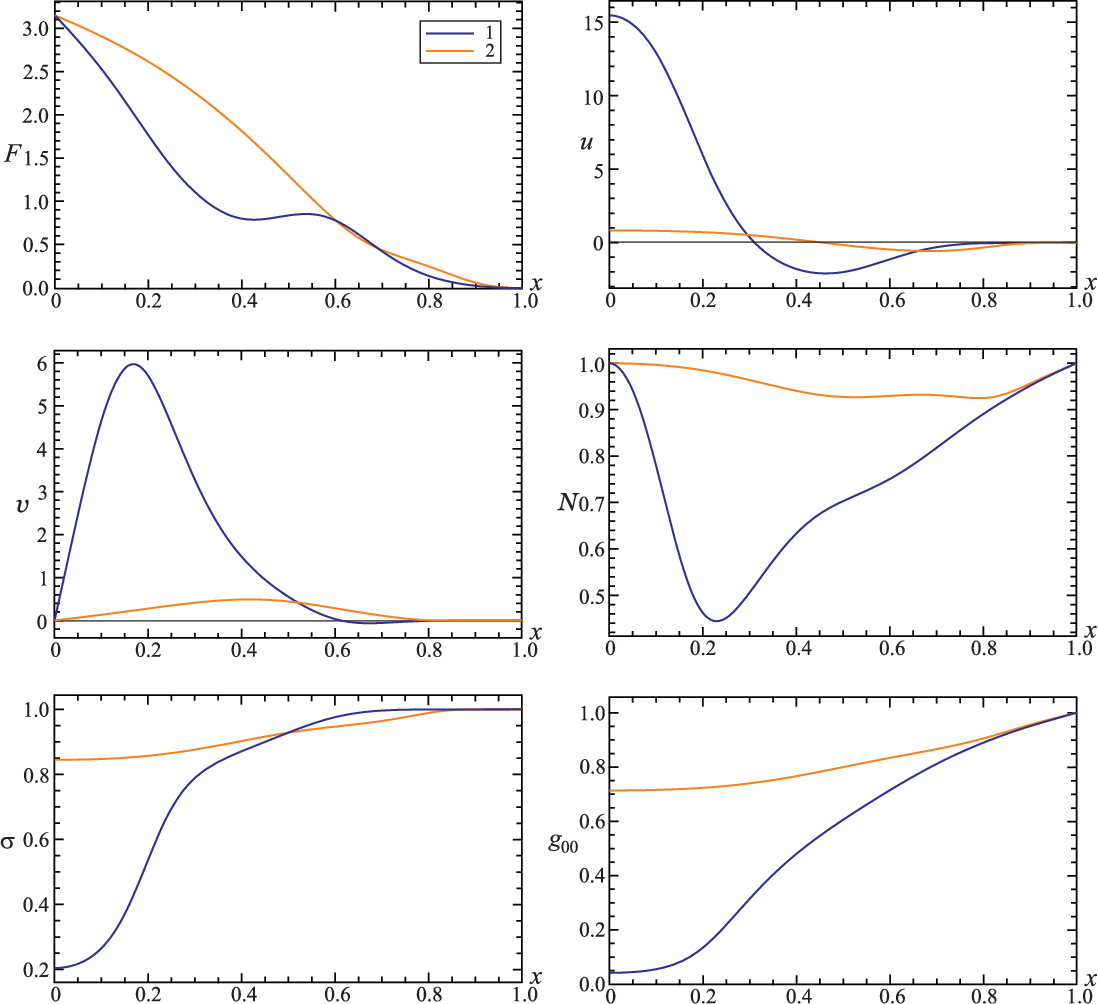}
        \vspace{-1.cm}
    \end{center}
    \caption{The profile functions $F$, $u$, $v$, $N$, and $\sigma$ 
    of the radially excited solutions 
    are shown versus the compactified radial coordinate $x$
    for the points 1 and 2 on the $\alpha=0.1$ loop
    in Fig.~\ref{fig_freq_coupling_one_node}.
     }
    \label{fig_field_distr_3}
\end{figure}

So far we here addressed only the fundamental spherically symmetric gravitating
Skyrmion-fermion solutions which have monotonic matter field functions 
$F(r)$, $u(r)$, and $v(r)$.
However, we find in addition discrete families of 
radially excited solutions that feature an increasing
number of zeros of their fermion functions. 
The corresponding spectral flow of these radially excited solutions 
is rather different from
the spectral flow of the nodeless solutions discussed above. 

As a particular example, we now consider
configurations with one node of the fermion functions $u$ and $v$ each, 
and a nodeless Skyrmion profile function,
restricting again to the case of vanishing bare mass, $m=0$.
We display in Fig.~\ref{fig_freq_coupling_one_node}
the scaled spectral flow for these radially excited solutions
for several values of the effective coupling $\alpha$.
Note that in contrast to the figures of the nodeless solutions,
where $\alpha^2$ was specified,
here $\alpha$ is shown in the legend.

Figure \ref{fig_freq_coupling_one_node}a 
exhibits the scaled eigenvalue $\omega/h$ versus the Yukawa coupling $h$,
while Fig.~\ref{fig_freq_coupling_one_node}b 
shows the scaled ADM mass $M/\alpha^2$ versus $h$. 
For small effective gravitational coupling $\alpha$, $\alpha^2 \le 0.04$, 
two branches of solutions emerge from the positive continuum 
at some critical value of the Yukawa coupling $h_\text{cr}<0$.
Interestingly, this value is close to the critical value $h_\text{max}$,
where the nodeless solutions emerge from the positive continuum
in Minkowski space (see Fig.~\ref{fig_freq_coupling}).

As seen in Fig.~\ref{fig_freq_coupling_one_node}a, 
the scaled eigenvalue $\omega/h$ of the localized fermionic mode
forms a typical double loop curve
consisting of two crossing branches,
when considered versus the Yukawa coupling $h$.
By analogy with our consideration above, 
we may possibly identify the evolution of the configuration 
along one of these branches as related to the decrease of the Yukawa coupling,
while the evolution along the second branch might be considered as being
obtained by decreasing the Skyrme parameter $f_\pi$.

To obtain a better understanding of this branch structure, 
we consider two sets of solutions,
labeled by points 1 and 2 in Fig.~\ref{fig_freq_coupling_one_node}a. 
We show their profile functions $F$, $u$, $v$, $N$, and $\sigma$ 
versus the compactified radial coordinate $x$
in Fig.~\ref{fig_field_distr_3}.
The solution with its fermionic mode 
closer to the positive continuum (point 2) is weakly localized. 
The backreaction of this mode on the Skyrme field is small, 
and the metric functions are not far from the flat space limit. 
On the contrary, 
the solution on the second branch (point 1) is strongly localized, 
the backreaction of the fermionic mode is very significant, 
and the gravitational interaction is strong.

Addressing now the associated ADM mass 
of these radially excited solutions,
the branch structure becomes more evident.
As seen in Fig.~\ref{fig_freq_coupling_one_node}b,
both branches bifurcate 
at a maximal value of the scaled ADM mass $M/\alpha^2$,
that corresponds to a minimal value of 
the Yukawa coupling $h$.
With increasing effective coupling $\alpha$,
the maximal value of the scaled ADM mass decreases,
while the minimal value of the Yukawa coupling increases. 
In fact, the solutions with one node rapidly degenerate, 
as the coupling to gravity becomes stronger,
and they cease to exist as $\alpha^2 > 0.04$. 
We note that we do not observe bifurcations 
between the solutions with one node 
and the nodeless configurations.

Most importantly, we note that none of the solutions
with one node exhibit a zero mode.
The double loop formed by the scaled eigenvalues $\omega/h$
never crosses zero, but remains always negative.
Consequently, their ADM mass neither becomes zero
or turns negative.
Those intriguing effects are only observed 
for the fundamental (nodeless) solutions.

\section{Conclusions}

Here we have considered gravitating solutions of Einstein-Skyrme-Dirac theory.
The bosonic sector of this theory gives rise to localized solutions,
connected to the Skyrmions in flat space 
and to the Bartnik-McKinnon solutions of Einstein-Yang-Mills theory.
In the presence of gravity these globally regular solutions
spawn hairy black hole solutions in General Relativity  \cite{Luckock:1986tr,Glendenning:1988qy,Droz:1991cx,Heusler:1991xx,Bizon:1992gb,Heusler:1993ci}.

Since Skyrmions are topological solitons,
the index theorem guarantees the occurrence of the spectral flow
of a normalizable fermionic mode bound to them.
In Minkowski space the backreaction of a spin-isospin singlet fermionic mode
on the spherically symmetric 
Skyrmion with unit topological charge was studied already long ago 
\cite{Hiller:1986ry,Kahana:1984be,Balachandran:1998zq}.
Depending on the Yukawa coupling strength,
this mode emerges from the positive continuum,
crosses zero and evolves toward the negative continuum.

Here we have presented a detailed study of the generalization 
of this phenomenon by minimally coupling the matter fields to gravity
\cite{Dzhunushaliev:2024iag}.
When gravity is added, 
we basically observe two branches of localized solutions,
the Skyrmion branch and the BM branch,
as the Yukawa coupling strength is varied,
instead of the single branch present in Minkowski space.
The two branches then emerge at a bifurcation point,
that is not connected to the positive continuum.

For small gravitational coupling the branches still emerge
at bifurcation points with positive eigenvalue of the fermionic mode, 
and then evolve toward smaller eigenvalues, 
with the modes on the Skyrmion branch
evolving toward the negative continuum.
For larger gravitational coupling the spectral flow
features merely negative eigenvalues of the fermionic mode.
Taking into account the backreaction of the fermionic mode
on the gravitating bosonic configurations 
thus implies dramatic consequences:
The ADM mass of the Skyrmion-fermion systems
crosses zero along the Skyrmion branches and then turns negative.
As one may expect, this is associated with the violation
of the energy conditions.

While we have mostly considered the backreacting Skyrmion-fermion systems
for vanishing bare mass of the fermions, 
we have also shown that the presence of a finite fermionic bare mass 
does not alter the basic scenario of the 
evolution of the spectral flow significantly.
In contrast, when we consider instead of the fundamental singlet mode
a radially excited state, the spectral flow changes drastically:
neither zero modes nor negative ADM masses arise.
 
It may be of interest to compare with another well-known system, 
the self-gravitating non-Abelian monopole \cite{Lee:1991vy,Breitenlohner:1991aa,Breitenlohner:1994di}.
In both cases the flat space configurations represent 
spherically symmetric topological solitons in (3+1)-dimensions, 
and there is a spin-isospin singlet fermionic mode localized on the soliton.

However, the pattern of the dynamical evolution of the gravitating monopole 
and of the Skyrmion is different.
While the gravitating monopole solutions bifurcate 
with an extremal Reissner-Nordstr\"{o}m black hole,
the gravitating Skyrmions bifurcate with a second branch of solutions,
tending to the regular scaled BM solution.
Moreover, the eigenvalue of the Dirac operator of the fermionic mode 
localized on the monopole is always zero.
It does not depend on the strength of the Yukawa coupling.
This mode is fully absorbed into the interior of the black hole 
as the configuration approaches the Reissner-Nordstr\"om limit 
\cite{Dzhunushaliev:2023ylf}.

We note that the possibility of the emergence of a negative mass 
in General Relativity is being discussed since a long time
\cite{Bondi:1957zz,Harari:1990cz} (for some recent work, see, e.g., 
Refs.~\cite{Mbarek:2014ppa,Farnes:2017gbf,Hao:2023kvf}).
It has also been pointed out that a region of negative energy density 
may collapse to a black hole 
with an unusual topology of the event horizon \cite{Mann:1997jb}.
One might therefore wonder 
whether, in the case of Skyrmion-fermion systems with negative mass,
the possibility of forming a black hole of negative mass 
carrying Skyrmion-fermion hair might arise.
This would be of particular interest, 
since previous attempts to obtain black holes with fermion hair 
in (3+1)-dimensional asymptotically flat spacetime have failed 
\cite{Finster:1998ws,Herdeiro:2019mbz,Herdeiro:2021jgc,Jing:2004xv}.

Finally, we would like to remark that it was suggested 
that Skyrmions might have technological significance in the future, 
providing fuel for the engines of Star Trek starships~\cite{Krusch:2004uf}.
Our present results indicate that Skyrmions 
might even have far wider implications, 
providing examples of anti-gravitating matter.

\section*{Acknowledgment}
We are grateful to Ioseph Buchbinder, Eugen Radu and Alexander Vikman  for inspiring and valuable discussions. Y.S. would like to thank the Hanse-Wissenschaftskolleg Delmenhorst for support and  hospitality. J.K. gratefully acknowledges support by DFG project Ku612/18-1.
This research was funded by the Committee of Science of the Ministry of Science and Higher Education of the Republic of Kazakhstan. 




\begin{thebibliography}{999}
\bibitem{tHooft:1974kcl}
G.~'t Hooft,
Nucl. Phys. B \textbf{79}, 276 (1974).
 \bibitem{Polyakov:1974ek}
A.~M.~Polyakov,
JETP Lett. \textbf{20}, 194 (1974).
 \bibitem{Skyrme:1961vq}
T.~H.~R.~Skyrme,
Proc. Roy. Soc. Lond. A \textbf{260}, 127 (1961).
\bibitem{Skyrme:1962vh}
T.~H.~R.~Skyrme,
Nucl. Phys. \textbf{31}, 556 (1962).
 \bibitem{Faddeev:1975tz}
L.~D.~Faddeev,
Print-75-0570 (IAS, PRINCETON).
 \bibitem{Faddeev:1996zj}
L.~D.~Faddeev and A.~J.~Niemi,
Nature \textbf{387}, 58 (1997).
\bibitem{Manton:2004tk}
N.~S.~Manton and P.~Sutcliffe, {\it Topological solitons} (Cambridge University Press, 2004).
\bibitem{Shnir:2018yzp}
Y.~M.~Shnir, {\it Topological and Non-Topological Solitons in Scalar Field Theories} (Cambridge University Press, 2018).
\bibitem{Volkov:1998cc}
M.~S.~Volkov and D.~V.~Gal'tsov,
Phys. Rept. \textbf{319}, 1 (1999).
\bibitem{Volkov:2016ehx}
M.~S.~Volkov,
\textit{``Hairy black holes in the XX-th and XXI-st centuries''}, Proceedings of the Fourteenth Marcel Grossmann Meeting, eds M.~Bianchi, R.~Jantzen and R.~Ruffini, World Scientific, pp. 1779-1798 (2017).
\bibitem{Lee:1991vy}
K.~M.~Lee, V.~P.~Nair, and E.~J.~Weinberg,
Phys. Rev. D \textbf{45}, 2751 (1992).
\bibitem{Breitenlohner:1991aa}
P.~Breitenlohner, P.~Forgacs, and D.~Maison,
Nucl. Phys. B \textbf{383}, 357 (1992).
\bibitem{Breitenlohner:1994di}
P.~Breitenlohner, P.~Forgacs, and D.~Maison,
Nucl. Phys. B \textbf{442}, 126 (1995).
\bibitem{Luckock:1986tr}
  H.~Luckock and I.~Moss,
  Phys.\ Lett.\ B {\bf 176}, 341 (1986).
\bibitem{Glendenning:1988qy}
N.~K.~Glendenning, T.~Kodama, and F.~R.~Klinkhamer,
Phys. Rev. D \textbf{38}, 3226 (1988).
\bibitem{Droz:1991cx}
  S.~Droz, M.~Heusler, and N.~Straumann,
  Phys.\ Lett.\ B {\bf 268}, 371 (1991).
\bibitem{Heusler:1991xx}
M.~Heusler, S.~Droz, and N.~Straumann,
Phys. Lett. B \textbf{271}, 61 (1991).
\bibitem{Bizon:1992gb}
P.~Bizon and T.~Chmaj,
Phys. Lett. B \textbf{297}, 55 (1992).
\bibitem{Heusler:1993ci}
M.~Heusler, N.~Straumann, and Z.~h.~Zhou,
Helv. Phys. Acta \textbf{66}, 614 (1993).
\bibitem{Atiyah:1975jf}
M.~F.~Atiyah, V.~K.~Patodi, and I.~M.~Singer,
Math. Proc. Cambridge Phil. Soc. \textbf{77}, 43 (1975).
\bibitem{Caroli:1964}
C.~Caroli, P.G.~de~Gennes, and J.~Matricon,
Phys. Lett.\textbf{9}, 307 (1964).
\bibitem{Jackiw:1975fn}
R.~Jackiw and C.~Rebbi,
Phys. Rev. D \textbf{13}, 3398 (1976).
\bibitem{Dashen:1974cj}
R.~F.~Dashen, B.~Hasslacher, and A.~Neveu,
Phys. Rev. D \textbf{10}, 4130 (1974).
\bibitem{Chu:2007xh}
Y.~Z.~Chu and T.~Vachaspati,
Phys. Rev. D \textbf{77}, 025006 (2008).
\bibitem{Nohl:1975jg}
C.~R.~Nohl,
Phys. Rev. D \textbf{12}, 1840 (1975).
\bibitem{Boguta:1985ut}
J.~Boguta and J.~Kunz,
Phys. Lett. B \textbf{154}, 407 (1985).
\bibitem{Callias:1977cc}
C.~J.~Callias,
Phys. Rev. D \textbf{16}, 3068 (1977).
\bibitem{Hiller:1986ry}
J.~R.~Hiller and T.~F.~Jordan,
Phys. Rev. D \textbf{34}, 1176 (1986).
\bibitem{Kahana:1984be}
S.~Kahana and G.~Ripka,
Nucl. Phys. A \textbf{429}, 462 (1984).
\bibitem{Balachandran:1998zq}
A.~P.~Balachandran and S.~Vaidya,
Int. J. Mod. Phys. A \textbf{14}, 445 (1999).
\bibitem{Stojkovic:2000ub}
D.~Stojkovic,
Phys. Rev. D \textbf{63}, 025010 (2001).
\bibitem{Jackiw:1981ee}
R.~Jackiw and P.~Rossi,
Nucl. Phys. B \textbf{190}, 681 (1981).
\bibitem{Rubakov:1982fp}
V.~A.~Rubakov,
Nucl. Phys. B \textbf{203}, 311 (1982).
\bibitem{Callan:1982au}
C.~G.~Callan, Jr.,
Phys. Rev. D \textbf{26}, 2058 (1982).
\bibitem{Witten:1984eb}
E.~Witten,
Nucl. Phys. B \textbf{249}, 557 (1985).
\bibitem{Gani:2010pv}
V.~A.~Gani, V.~G.~Ksenzov, and A.~E.~Kudryavtsev,
Phys. Atom. Nucl. \textbf{73}, 1889 (2010).
\bibitem{Amado:2014waa}
A.~Amado and A.~Mohammadi,
Eur. Phys. J. C \textbf{77}, 465 (2017).
\bibitem{Klimashonok:2019iya}
V.~Klimashonok, I.~Perapechka, and Y.~Shnir,
Phys. Rev. D \textbf{100}, 105003 (2019).
\bibitem{Perapechka:2018yux}
I.~Perapechka, N.~Sawado, and Y.~Shnir,
JHEP \textbf{10}, 081 (2018).
\bibitem{Amari:2024rpm}
Y.~Amari, N.~Sawado, and S.~Yamamoto,
JHEP \textbf{06} (2024), 057
\bibitem{Perapechka:2019upv}
I.~Perapechka and Y.~Shnir,
Phys. Rev. D \textbf{99}, 125001 (2019).
\bibitem{Perapechka:2019vqv}
I.~Perapechka and Y.~Shnir,
Phys. Rev. D \textbf{101}, 021701 (2020).
\bibitem{Campos:2022flw}
J.~G.~F.~Campos and A.~Mohammadi,
JHEP \textbf{08}, 180 (2022).
\bibitem{Gani:2022ity}
V.~A.~Gani, A.~Gorina, I.~Perapechka, and Y.~Shnir,
Eur. Phys. J. C \textbf{82}, 757 (2022).
\bibitem{Bazeia:2022yyv}
D.~Bazeia, J.~G.~F.~Campos, and A.~Mohammadi,
JHEP \textbf{12}, 085 (2022).
\bibitem{Saadatmand:2022htx}
D.~Saadatmand and H.~Weigel,
Phys. Rev. D \textbf{107}, 036006 (2023).
\bibitem{Weigel:2023fxe}
H.~Weigel and D.~Saadatmand,
Universe \textbf{10}, 13 (2024).
\bibitem{Taub:1937zz}
A.~H.~Taub,
Phys. Rev. \textbf{51}, 512 (1937).
\bibitem{Volkov:1994tp}
M.~S.~Volkov,
Phys. Lett. B \textbf{334}, 40 (1994).
\bibitem{Kunz:1993ir}
J.~Kunz and Y.~Brihaye,
Phys. Lett. B \textbf{304}, 141 (1993)
\bibitem{Dzhunushaliev:2023ylf}
V.~Dzhunushaliev, V.~Folomeev, and Y.~Shnir,
Phys. Rev. D \textbf{108}, 065005 (2023).
\bibitem{Finster:1998ws}
F.~Finster, J.~Smoller, and S.~T.~Yau,
Phys. Rev. D \textbf{59}, 104020 (1999).
\bibitem{Herdeiro:2019mbz}
C.~Herdeiro, I.~Perapechka, E.~Radu, and Y.~Shnir,
Phys. Lett. B \textbf{797}, 134845 (2019).
\bibitem{Dzhunushaliev:2018jhj}
V.~Dzhunushaliev and V.~Folomeev,
  Phys.\ Rev.\ D {\bf 99}, 084030 (2019).
\bibitem{Dzhunushaliev:2019kiy}
V.~Dzhunushaliev and V.~Folomeev,
  Phys.\ Rev.\ D {\bf 99}, 104066 (2019).
\bibitem{Herdeiro:2021jgc}
C.~Herdeiro, I.~Perapechka, E.~Radu, and Y.~Shnir,
Phys. Lett. B \textbf{824}, 136811 (2022).
\bibitem{Blazquez-Salcedo:2019uqq}
J.~L.~Bl\'azquez-Salcedo and C.~Knoll,
Eur. Phys. J. C \textbf{80}, 174 (2020).
\bibitem{Blazquez-Salcedo:2020czn}
J.~L.~Bl\'azquez-Salcedo, C.~Knoll, and E.~Radu,
Phys. Rev. Lett. \textbf{126}, 101102 (2021).
\bibitem{Bolokhov:2021fil}
S.~Bolokhov, K.~Bronnikov, S.~Krasnikov, and M.~Skvortsova,
Grav. Cosmol. \textbf{27}, 401 (2021).
\bibitem{Konoplya:2021hsm}
R.~A.~Konoplya and A.~Zhidenko,
Phys. Rev. Lett. \textbf{128}, 091104 (2022).
\bibitem{Armendariz-Picon:2003wfx}
C.~Armendariz-Picon and P.~B.~Greene,
Gen. Rel. Grav. \textbf{35}, 1637 (2003).
\bibitem{Cai:2008gk}
Y.~F.~Cai and J.~Wang,
Class. Quant. Grav. \textbf{25}, 165014 (2008).
\bibitem{Dzhunushaliev:2024iag}
V.~Dzhunushaliev, V.~Folomeev, J.~Kunz, and Y.~Shnir,
Phys. Lett. B \textbf{855}, 138812 (2024).
\bibitem{Bartnik:1988am}
R.~Bartnik and J.~Mckinnon,
Phys. Rev. Lett. \textbf{61}, 141 (1988).
\bibitem{Dolan:2015eua}
S.~R.~Dolan and D.~Dempsey,
Class. Quant. Grav. \textbf{32}, 184001 (2015).
\bibitem{Eguchi:1980jx}
T.~Eguchi, P.~B.~Gilkey, and A.~J.~Hanson,
Phys. Rept. \textbf{66}, 213 (1980).
\bibitem{Gell-Mann:1960mvl}
M.~Gell-Mann and M.~Levy,
Nuovo Cim. \textbf{16}, 705 (1960).
\bibitem{Krusch:2003xh}
S.~Krusch,
J. Phys. A \textbf{36}, 8141 (2003).
\bibitem{Shnir:2002dw}
Y.~Shnir,
Phys. Scripta \textbf{67}, 361 (2003).
\bibitem{Jackiw:1976xx}
R.~Jackiw and C.~Rebbi,
Phys. Rev. Lett. \textbf{36}, 1116 (1976).
\bibitem{Adkins:1983ya}
G.~S.~Adkins, C.~R.~Nappi, and E.~Witten,
Nucl. Phys. B \textbf{228}, 552 (1983).
\bibitem{Manton:2006tq}
N.~S.~Manton and S.~W.~Wood,
Phys. Rev. D \textbf{74}, 125017 (2006).
\bibitem{pardiso}N.I.M.~Gould, J.A.~Scott, Y.~Hu,
ACM Trans. Math. Softw. {\bf 33}, 10 (2007);\\
O.~Schenk, K.~Gartner, 
Future Gener. Comput. Syst. {\bf 20}, 475 (2004).
\bibitem{Kunz:2019sgn}
J.~Kunz, I.~Perapechka, and Y.~Shnir,
JHEP \textbf{07}, 109 (2019).
\bibitem{Rubakov:2014jja}
V.~A.~Rubakov,
Phys. Usp. \textbf{57}, 128 (2014).
\bibitem{Bondi:1957zz}
H.~Bondi,
Rev. Mod. Phys. \textbf{29}, 423 (1957).
\bibitem{Harari:1990cz}
D.~Harari and C.~Lousto,
Phys. Rev. D \textbf{42}, 2626 (1990).
\bibitem{Mbarek:2014ppa}
S.~Mbarek and M.~B.~Paranjape,
Phys. Rev. D \textbf{90}, 101502 (2014).
\bibitem{Farnes:2017gbf}
J.~S.~Farnes,
Astron. Astrophys. \textbf{620}, A92 (2018).
\bibitem{Hao:2023kvf}
C.~H.~Hao, L.~X.~Huang, X.~Su, and Y.~Q.~Wang,
[arXiv:2312.03800 [gr-qc]].
\bibitem{Mann:1997jb}
R.~B.~Mann,
Class. Quant. Grav. \textbf{14}, 2927 (1997).
\bibitem{Jing:2004xv}
J.~l.~Jing,
Phys. Rev. D \textbf{70}, 065004 (2004).
\bibitem{Krusch:2004uf}
S.~Krusch and P.~Sutcliffe,
J. Phys. A \textbf{37}, 9037 (2004).
\end{thebibliography}
 \end{document}